\documentclass[10pt]{article}
\usepackage{euscript,amssymb,epsfig}
\catcode`\@=11

\newcommand{\be}[1]{\begin{equation}\label{#1}}
\newcommand{\ee}{\end{equation}}
\newcommand{\ba}[1]{\begin{eqnarray}\label{#1}}
\newcommand{\ea}{\end{eqnarray}}
\newcommand{\rf}[1]{(\ref{#1})}
\newcommand{\nn}{\nonumber}
\newcommand{\ov}{\overline}

\newcommand{\td}{\tilde}
\newcommand{\etal}{{\it et al }}
\newcommand{\ibid}{{\it ibid. }}
\newcommand{\bmatrix}[1]{\left( \begin{array}{#1}}
\newcommand{\ematrix}{\end{array}\right)}
\newcommand{\opensquare}{\mbox{$\rlap{$\sqcap$}\sqcup$}}

\newcommand{\const}{\mbox{\rm const}\,}

\newcommand{\R}{ \mbox{\rm I$\!$R} }
\newcommand{\sign}{ \mbox{\rm sign}\,}

\font\msbm=msbm10
\def\RR{\hbox{\msbm R}}

\title{\mbox{}\hfill {\large
AEI-2003-015}
\\[1.5cm]
Nonlinear multidimensional cosmological models with form fields:
stabilization of extra dimensions and the cosmological constant
problem}
\author{U. G\"unther$^{ad}$\footnote{e-mail:
u.guenther@fz-rossendorf.de}~\footnote{present address:
 Research Center Rossendorf, P.O. Box 510119, D-01314 Dresden, Germany}\, ,
P. Moniz$^b$\footnote{e-mail: pmoniz@dfisica.ubi.pt}~
\footnote{Also at Centra --- IST, Rua Rovisco Pais, 1049 Lisboa,
Portugal} \, and A. Zhuk$^{cd}$ \footnote{e-mail:
zhuk@paco.net}\\[2ex]
$^a$ Gravitationsprojekt, Mathematische
Physik I,\\ Institut f\"ur Mathematik, Universit\"at Potsdam,\\ Am
Neuen Palais 10, PF 601553, D-14415 Potsdam, Germany
\\[1ex]
$^b$ Departamento de F$\acute{\mbox{\i}}$sica,
Universidade da Beira Interior,\\ Rua Marqu$\hat{e}$s
D'$\acute{A}$vila e Bolama, 6200 Covilh$\tilde{a}$, Portugal
\\[1ex]
  $^c$ Department of Physics, University of Odessa,\\ 2
Dvoryanskaya St., Odessa 65100, Ukraine\\[1ex]
$^d$ Max-Planck-Institut f\"ur Gravitationsphysik,\\
Albert-Einstein-Institut,\\
Am M\"uhlenberg 1,
D-14476 Golm, Germany\\[1ex] }

\date{02 March 2003}
\begin{document}

\maketitle
\begin{abstract}
We consider multidimensional gravitational models with a nonlinear
scalar curvature term and  form fields in the action functional.
In our scenario it is assumed that   the higher dimensional
spacetime undergoes a spontaneous compactification to a warped
product manifold. Particular attention is paid to models with
quadratic scalar curvature terms and a Freund-Rubin-like ansatz
for solitonic form fields. It is shown that for certain parameter
ranges the extra dimensions are stabilized. In particular,
stabilization is possible for any sign of the internal
 space curvature,   the bulk cosmological constant and of the
effective four-dimensional cosmological constant.  Moreover, the
effective cosmological constant can satisfy the observable limit
on the dark energy density. Finally, we discuss the restrictions
on the parameters of the considered nonlinear models and how they
 follow from the connection between the D--dimensional and the
four-dimensional fundamental
 mass scales.
\end{abstract}


\vspace{.5cm}

\section{Introduction}
\setcounter{equation}{0}

Two of the most intriguing problems of modern cosmology are the
problem of additional dimensions and the cosmological constant
problem (CCP). The first problem follows from theories which unify
different fundamental interactions with gravity, such as M/string
theory \cite{pol-wit}, and which have their most consistent
formulation in spacetimes with more than four dimensions. The
problem can be naturally formulated  as the following question: if
we live in a multidimensional spacetime, why do we not observe the
extra dimensions?  Within the "old" Kaluza-Klein (KK) framework
and the early $E_8\times E_8$-heterotic string phenomenology the
question is answered by assuming the extra dimensions so small
(i.e. with a characteristic size $r$ between the Planck and the
Fermi scales $10^{-33}$cm $\lesssim r \lesssim 10^{-17}$cm) that
they are not accessable by present-day collider experiments. New
concepts with the possibility for rich phenomenology opened up
with the uncovering of $D-$branes by Polchinski \cite{pol-1} in
1995. In "brane-world" scenarios of the Universe  the usual
$4-$dimensional physics with its $SU(3)\times SU(2)\times U(1)$
standard model (SM) fields is localized on a $3-$dimensional
space-like hypersurface (our world-brane) whereas the
gravitational field propagates in the whole (bulk) spacetime.
Depending on the concrete scenario there are different types of
masking of the additional dimensions. Whereas in
Arkani-Hamed--Dvali--Dimopoulos (ADD) models
\cite{sub-mill1,sub-mill1a,sub-mill2} the extra dimensions are
curled up to sizes smaller than $10^{-2}$cm, so that they are in
agreement with present table-top Cavendish-type tests of gravity
\cite{experiment1}, they can be infinite
 in the Randall-Sundrum II (RS II) \cite{RSII} and the
Dvali--Gabadadze--Porrati (DGP) \cite{DGP} model. In the latter
models the appearing four-dimensionality of low-energy physics is
achieved by inducing appropriate effective gravitational potentials
 on the world-brane.  Beside their interesting phenomenology, brane-world
models provide a possible resolution of the hierarchy problem. In
ADD-type models this is due to the connection between the
 Planck scale $M_{Pl(4)}$ and the fundamental scale
$M_{*(4+D^{\prime})}$ of the $4-$dimensional and the
(4+$D^{\prime}$)-dimensional spacetime, respectively:
\be{0.1}
M_{Pl(4)}^2 \sim V_{D^{\prime}}
M_{*(4+D^{\prime})}^{2+D^{\prime}}\, .
\ee
$V_{D^{\prime}}$ denotes the volume of the compactified
$D^{\prime}$ extra dimensions. It was realized in
\cite{sub-mill1,sub-mill1a,sub-mill2} that  localizing the SM
fields on a $3-$brane allows to lower $M_{*(4+D^{\prime})}$ down
to the electroweak scale $M_{EW} \sim 1$TeV without contradiction
with present observations. Therefore, the compactification scale
of the internal space can be of order
\be{0.2} r \sim V_{D^{\prime}}^{1/D^{\prime}} \sim
10^{\frac{32}{D^{\prime}}-17} \mbox{cm}\, . \ee
With $M_{EW} \sim 1$TeV, physically acceptable values correspond
to $D^{\prime}\ge 3$ \cite{sub-mill1} (for astrophysical and
cosmological bounds see e.g. \cite{sub-mill2a}; experimental
bounds from table-top Cavendish-type experiments are given in
\cite{experiment1}), and for $D^{\prime} =3$ one arrives at a
sub-millimeter compactification scale $r\sim 10^{-5} \mbox{mm}$ of
the internal space. If we shift $M_{*(4+D^{\prime})}$ to $30$TeV,
as suggested in \cite{sub-mill2a} (see also \cite{Rubakov}), then
the $D^{\prime} =2$ case satisfies all aforementioned bounds and
leads to  $r\sim 10^{-3}-10^{-2} \mbox{mm}$. In order to not
exclude this $D^{\prime} =2$ with its largest possible
compactification scale $r$, we assume that the fundamental scale
$M_{*(4+D^{\prime})}$ can be of order $30$TeV. Additionally, the
geometry in the ADD approach is assumed to be factorizable as in a
standard Kaluza-Klein  model. I.e., the topology is the direct
product of a non-warped external spacetime manifold and internal
space manifolds with warp factors which depend on the external
coordinates\footnote{The M-theory inspired RS-scenarios
\cite{RSII,RSI} use a non-factorizable geometry with $D^{\prime}
=1$. Here, the $4-$dimensional spacetime is warped with a factor
$\tilde\Omega$ which depends on the extra dimension and Eq.
\rf{0.1} is modified as: $M_{Pl(4)} \sim \tilde\Omega^{-1}M_{EW}$.
In our paper we concentrate on the factorizable geometry of
ADD-type models.\label{RS}}.

According to observations, the internal space should be static or
nearly static at least from the time of primordial
nucleosynthesis, (otherwise the fundamental physical constants
would vary, see e.g. \cite{GZ(CQG2),CV}). This means that at the
present evolutionary stage of the Universe  the compactification
scale of the internal space should either be stabilized and
trapped at the minimum of some effective potential, or it should
be slowly varying (similar to the slowly varying cosmological
constant in the quintessence scenario \cite{WCOS}). In both cases,
small fluctuations over stabilized or slowly varying
compactification scales (conformal scales/geometrical moduli) are
possible.

Stabilization of extra dimensions (moduli stabilization) in models
with large extra dimensions (ADD-type models) has been considered
in a number of papers (see e.g., Refs.
\cite{sub-mill2,d2,sub-mill3,CGHW,Geddes,demir,NSST,PS})\footnote{In
most of these papers, moduli stabilization was  considered without
regard to the energy-momentum localized on the brane so that the
dynamics of the multidimensional universe was mainly defined by
the energy-momentum of the bulk matter. A brane matter
contribution was taken into account, e.g., in \cite{PS}.}. In the
corresponding approaches, a product topology of the $(4+D^{\prime
})-$dimensional bulk spacetime was constructed from Einstein
spaces with scale (warp) factors depending only on the coordinates
of the external $4-$dimensional component. As a consequence, the
conformal excitations had the form of massive scalar fields living
in the external spacetime. Within the framework of
multidimensional cosmological models (MCM) such excitations were
investigated in \cite{GZ1,GZ,GZ(PRD2)} where they were called
gravitational excitons. Later, since the ADD
compactification approach these geometrical moduli excitations are
 known as radions \cite{sub-mill2,sub-mill3}.

Most of the aforementioned  papers are devoted to the
stabilization of large extra dimension in theories with a linear
multidimensional gravitational action. String theory suggests that
the usual linear Einstein-Hilbert action should be extended with
higher order nonlinear curvature terms. In a previous paper
\cite{GMZ(PRDa)} we considered a simplified model with
multidimensional Lagrangian of the form $L = f(R)$, where $f(R)$
is an arbitrary smooth function of the scalar curvature. Without
connection to stabilization of the extra-dimensions, such models
($4-$dimensional as well as multi-dimensional ones) were
considered  e.g. in Refs. \cite{Kerner,Maeda,EKOY}. There, it was
shown that the nonlinear models are equivalent to models with
linear gravitational action plus a minimally coupled scalar field
with self-interaction potential. In \cite{GMZ(PRDa)}, we advanced
this equivalence towards investigating the stabilization problem
for extra dimensions. Particular attention was paid to models with
quadratic scalar curvature terms. It was shown that for certain
parameter ranges, the extra dimensions are stabilized if the
internal spaces have negative constant curvature. In this case,
the 4--dimensional effective cosmological constant $\Lambda_{eff}$
as well as the bulk cosmological constant $\Lambda_D$ become
negative. As a consequence, the homogeneous and isotropic external
space is asymptotically $\mbox{AdS}_4$. Because the considered
nonlinear model is a pure geometrical one (only with a bare
cosmological constant $\Lambda_D$ as an exotic matter source
included) the equivalent linear model contains only a minimally
coupled scalar field as bulk matter. The null energy condition
(NEC) $T_{ab}N^aN^b \ge 0$ for this field reads $T_{ab}N^aN^b =
(N^a\partial_a\phi)^2 \ge 0$ (with $N$ a future directed null
vector) and is satisfied only marginally when the internal spaces
are completely stabilized and the scalar field is frozen out.
Moreover, the weak energy condition (WEC) $T_{ab}W^aW^b \ge 0$
(with $W$ a future directed time-like vector) is violated in this
case because the energy density $\rho $ of the scalar field is
negative definite $\rho <0$. As a result, the aforementioned
parameters (the internal space scalar curvatures, $\Lambda_D$ and
$\Lambda_{eff}$) are negative in the case of stabilized internal
spaces (see also \cite{CGHW,NSST,GZ(PRD2)}).

However, a negative cosmological constant leads to a deceleration
of the Universe instead to an accelerated expansion, as recent
observational data indicate. According to these data our Universe
 is dominated by a dark energy component with
negative pressure. For example, from observations of the clusters
of galaxies follows that the energy density of the matter
components which can clump in various structures is significantly
undercritical. But, the position of the first acoustic peak in the
angular power spectrum of the cosmic microwave background
radiation (CMB) implies that the Universe is, on large scales,
nearly flat. In other words, the energy density in the Universe is
very close to the critical value. Thus, there must exist a
homogeneously distributed exotic (dark) energy component
\cite{PR1}. This observation is in agreement with the conclusion
following from the Hubble diagram of type Ia supernovae (SN-Ia) at
high redshifts, which also indicate that our Universe  currently
undergoes an accelerated expansion. Under the assumption of
flatness, using the data of the  CMB anisotropy measurements, high
redshift SN-Ia observations and from local cluster abundances, the
authors of Ref. \cite{eq.state} found a constraint on the equation
of state parameter $\omega_Q = P/\rho < -0.85$ at 68\% of
confidence level. They concluded that this result is in perfect
agreement with the $\omega_Q =-1$ cosmological constant case and
gives no support to a quintessential field scenario with $\omega_Q
>-1$. Results obtained in \cite{SRSS} also favor $\omega_Q \approx
-1$ at the present epoch.

In Ref. \cite{GMZ(PRDa)} we already indicated that the effective
cosmological constant can be shifted from negative values to
positive ones by including into the nonlinear model matter fields
which satisfy the NEC. In the present paper, we demonstrate this
effect explicitly by endowing the extra dimensions with
real-valued solitonic form fields \cite{Stelle}. Such fields
naturally arise as Ramond-Ramond (RR) form fields in type II
string theory
 and  M-theory. Within a generalized Freund-Rubin
setting \cite{FR} their influence on the evolutionary dynamics of
the Universe has been considered, e.g., in Refs.
\cite{wiltshire,LOW,IM1a,GC1} and due to its simplicity we adopt
this ansatz here for the stability analysis of our nonlinear
model. {}From Eqs. \rf{1.6}, \rf{1.7} below, it can be easily seen
that the real-valued form fields satisfy the NEC as well as the
WEC. However, the strong energy condition (SEC) is violated in our
model by the cosmological constant\footnote{For a critical
discussion of the different ECs we refer to \cite{BV1}.}. The
presence of two types of fields in our equivalent linear model ---
the minimally coupled scalar field (which satisfies the NEC only
marginally and which can violate the WEC) and the form fields
(which satisfy both of these conditions)
---  leads to a rich and interesting picture of stable
configurations with various sign combinations for the allowed
cosmological constants as well as for the constant curvatures of
the internal space. Beside stability regions with negative
4--dimensional effective cosmological constant $\Lambda_{eff}<0$
the parameter space contains also regions with $\Lambda_{eff}>0$
which can ensure an accelerated expansion of the Universe.

As mentioned at the very beginning of the Introduction, there
still remains the problem about the incredible smallness of the
cosmological constant \cite{dolgov,ccp1}. Moreover, it is
completely unclear why its energy density is comparable with the
energy density of matter just at the present time (the cosmic
coincidence problem). Modern reviews on the cosmological constant
problem can be found for example in \cite{PR1,SahStar,Straumann}.
In our paper we show that for stabilized internal spaces a small
positive cosmological constant of the external (our) space can
arise from ADD- and KK-type multidimensional models. We
demonstrate that the smallness of the effective cosmological
constant can follow from a natural parameter choice of the
considered nonlinear ansatz. Unfortunately, the extremely small
value of the observed cosmological constant
 requires a very strong
fine tuning of the parameters.

The paper is structured as follows. The general setup of our model
is given in section \ref{setup}. In section \ref{compact}, we make
the geometry of the spacetime manifold explicit --- endowing the
internal space with the structure of a warped product of $n$
factor spaces (due to spontaneous compactification). Furthermore,
we specify the generalized Freund-Rubin ansatz for the form fields
and perform a dimensional reduction of the action functional to a
4-dimensional effective theory with $(n+1)$ self-interacting
minimally coupled scalar fields (section \ref{compact}). The
stabilization of the extra dimensions is then reduced to the
condition that the obtained effective potential for these fields
should have a minimum. In section \ref{stability}, we present a
detailed analysis of this problem for a model with one internal
space. The main results are summarized and discussed in the
concluding section \ref{conclu}.


\section{General setup\label{setup}}
\setcounter{equation}{0}

We consider a $D=(4+D')$ - dimensional nonlinear gravitational
theory with action
\be{1.1} S = \frac {1}{2\kappa_D^2}\int_M d^Dx \sqrt{|\ov g|}
f(\ov R) -\frac12 \int_M d^Dx \sqrt{|\ov g|}\sum_{i=1}^n
\frac{1}{d_i!} \left(F^{(i)} \right)^2\; , \ee
where $f(\ov R)$ is an arbitrary smooth function with mass
dimension $\mathcal{O}(m^2)$ \ ($m$ has the unit of mass) of the
scalar curvature $\ov R = R[\ov g]$ constructed from the
D--dimensional metric $\ov g_{ab}\; (a,b = 1,\ldots,D)$.
\be{1.1a}  \kappa^2_D = 8\pi /
M_{*(4+D^{\prime})}^{2+D^{\prime}}\ee
denotes the D--dimensional gravitational constant (subsequently,
we
 assume that $M_{*(4+D^{\prime})}\sim M_{EW}$). In action
 \rf{1.1},
$F^{(i)} = F^{(i)}_{m_i n_i \dots q_i}\, , \quad i=1, \dots ,n$
 is an
antisymmetric tensor field of rank $d_i$ (a $d_i-$form field
strength) with indices from an index set $s_{(i)}=\{m_i: \ \max
(m_i) - \min (m_i) =d_i\}$, where $ m_i, n_i, \ldots , q_i \in
s_{(i)}$.
 For simplicity, we suppose that the index sets $s_{(i)}$, $s_{(j)}$
of tensors $F^{(i)}$, $F^{(j)}$ with $i\neq j$  contain no common
elements as well as no indices corresponding to the coordinates of
the $D_0-$dimensional external spacetime (usually $D_0 =4$).
Additionally, we assume that for the sum of the ranks holds
$\sum_{i=1}^n d_i = D-D_0 := D^{\prime}$. Obviously, this model
can be generalized to tensor configurations $F^{(i)}$, $F^{(j)}$
with intersecting (overlapping) index sets. In this case explicit
field configuration can be obtained, e.g., when the indices
satisfy special overlapping rules \cite{IM1a}. Such a
generalization is beyond the scope of the present paper.
Furthermore, we assume in our subsequent considerations that the
index sets $m_i,n_i,\dots ,q_i \ne 0$ do not contain the
coordinates of the external spacetime $M_0$ and, hence, the field
strengths $F^{(i)}$ can be associated with a magnetic (solitonic)
$p-$brane system
 located in the extra dimensions as
discussed, e.g., in Refs. \cite{Stelle,LOW,IM1a}.

The equation of motion for the gravitational sector of \rf{1.1}
reads
\be{1.2} f^{\prime }\ov R_{ab} -\frac12 f \ov g_{ab} - \ov
\nabla_a \ov \nabla_b f^{\prime } + \ov g_{ab} \ov{\opensquare }
f^{\prime } = \kappa_D^2T_{ab} \left[ F,\ov g\right] \; , \ee
where $a,b = 1,\ldots,D$, $\; f^{\prime } =df/d\ov R$, $\; \ov
R_{ab} = R_{ab}[\ov g]$, $\; \ov R = R[\ov g]$. $\; \ov \nabla_a$
and $\ov{\opensquare}$ denote the covariant derivative and the
Laplacian with respect to the metric $\ov g_{ab}$
\be{1.3} \ov{\opensquare} = \opensquare [\ov g] = \ov g^{ab}\ov
\nabla_a \ov \nabla_b = \frac{1}{\sqrt{|\ov g|}}
\partial_a \left( \sqrt{|\ov g|}\quad \ov g^{ab}
\partial_b \right)\; .
\ee
Eq. \rf{1.2} can be rewritten in the form
\be{1.4} f^{\prime }\ov G_{ab} +\frac12 \ov g_{ab} \left( \ov R
f^{\prime} - f\right) - \ov \nabla_a \ov \nabla_b f^{\prime } +
\ov g_{ab} \ov{\opensquare } f^{\prime } = \kappa_D^2T_{ab} \left[
F,\ov g\right]\; , \ee
where $\ov G_{ab} = \ov R_{ab} -\frac12 \ov R \; \ov g_{ab}$, and
its trace
\be{1.5} (D-1)\ov{\opensquare } f^{\prime } = \frac{D}{2} f
-f^{\prime }\ov R + \kappa_D^2T\left[ F,\ov g\right]\;  \ee
 can be considered as a connection between $\ov R$ and $f$. The
energy momentum tensor (EMT) $T_{ab}\left[ F,\ov g\right]$ is
defined in the standard way as
\be{1.6} T_{ab}\left[ F,\ov g\right] \equiv \frac1{\sqrt{|\ov
g|}}\frac{\delta \left(\sqrt{|\ov g|} \sum_{i=1}^n
\frac{1}{d_i!}\left(F^{(i)} \right)^2\right)} {\delta \ov g^{ab}}
= \sum_{i=1}^n T_{ab}\left[ F^{(i)},\ov g\right]\, , \ee
where \be{1.7} T_{ab}\left[ F^{(i)},\ov g\right] =
\frac1{d_i!}\left(-\frac12\ov g_{ab} F^{(i)}_{m_i n_i \dots
q_i}F^{(i)\, m_i n_i \dots q_i} + d_i F^{(i)}_{a n_i\dots
q_i}F_b^{(i)\, n_i\dots q_i}\right)\, . \ee
For the trace of this tensor we obtain
\be{1.8} T\left[ F,\ov g\right] = \sum_{i=1}^n T\left[ F^{(i)},\ov
g\right] \ee
with
\be{1.9} T\left[ F^{(i)},\ov g\right] =
\frac{2d_i-D}{2(d_i!)}F^{(i)}_{m_i n_i\dots q_i} F^{(i)\, m_i
n_i\dots q_i}\, . \ee
The field strengths  $F^{(i)}$ satisfy the equations of motion
\be{1.10} {F^{(i)\, m_i n_i \dots q_i}}_{;\, q_i} = 0
\Longleftrightarrow \frac1{\sqrt{|\ov g|}} \left( \sqrt{|\ov g|}\;
F^{(i)\, m_i n_i \dots q_i}\right)_{,\, q_i}=0\, , \quad i=1,\dots
,n\, . \ee
and the Bianchi identities
\be{1.11} F^{(i)}_{[m_i n_i \dots q_i ,\; a]} = 0\, , \quad
i=1,\dots ,n\, . \ee

Following Refs. \cite{Kerner,Maeda,EKOY}, we perform a conformal
transformation
\be{1.12} g_{ab} = \Omega^2 \ov g_{ab}\;  \ee
with
\be{1.13} \Omega = \left[ f'(\ov R)\right]^{1/(D-2)}\;  \ee
and reduce the nonlinear gravitational theory to a linear one with
additional scalar field. This transformation is well defined for
$f'(\ov R)>0$ (concerning the case $f' \le 0$ see footnote
\ref{f'}). The equivalence of the theories can be easily proven
with the help of the auxiliary formulae
\be{1.14} \opensquare = \Omega^{-2}\left[\ov{\opensquare } +
(D-2)\ov g^{ab}\Omega^{-1}\Omega_{,a}\partial_b\right]
\Longleftrightarrow \ov {\opensquare} =
\Omega^{2}\left[\opensquare - (D-2) g^{ab}\Omega^{-1} \;
\Omega_{,a}\partial_b \right]\; , \ee
\be{1.15} R_{ab} = \ov R_{ab} +\frac{D-1}{D-2}( f')^{-2} \ov
{\nabla}_a f' \ov {\nabla}_b f' -(f')^{-1} \ov {\nabla}_a \ov
{\nabla}_b f'- \frac{1}{D-2} \ov g_{ab} (f')^{-1} \ov
{\opensquare} f' \ee
and
\be{1.16} R = (f')^{2/(2-D)}\left\{ \ov R +\frac{D-1}{D-2}
(f')^{-2} \ov g^{ab}\partial_a f'\partial_b f'-
2\frac{D-1}{D-2}(f')^{-1} \ov {\opensquare} f'\right\}\; . \ee
Defining the scalar $\phi $ by the relation
\be{1.19} f' = \frac {df}{d \ov R}
:= e^{A \phi} > 0\; ,\quad A := \sqrt{\frac{D-2}{D-1}}\; \ee and
making use of \rf{1.14} - \rf{1.16}, Eqs. \rf{1.4} and \rf{1.5}
can be rewritten as
\be{1.17} G_{ab} = \kappa^2_D T_{ab}\left[ F,\phi, g\right] +
T_{ab}\left[ \phi ,g\right] \ee
and
\be{1.18} \opensquare \phi = \frac {1}{\sqrt{(D-1)(D-2)}}\;
e^{\frac {-D}{\sqrt{(D-1)(D-2)}}\phi} \left( \frac {D}{2} f -
f'\ov R \right) + \frac1{\sqrt{(D-1)(D-2)}}\kappa^2_D T\left[
F,\phi ,g\right] \; . \ee
The EMTs read
\be{1.20} T_{ab}\left[ \phi,g\right] = \phi_{,a}\phi_{,b} -\frac12
g_{ab}g^{mn}\phi_{,m}\phi_{,n} - \frac12 g_{ab}\; e^{\frac
{-D}{\sqrt{(D-1)(D-2)}}\phi} \left(\ov R f'- f\right)\, , \ee
\be{1.21} T_{ab}\left[ F,\phi ,g\right] = \sum_{i=1}^n
e^{\frac{2d_i-D}{\sqrt{(D-1)(D-2)}}\phi} T_{ab}\left[
F^{(i)},g\right]
\ee
and
\be{1.22} T\left[ F,\phi,g\right] = \sum_{i=1}^n
e^{\frac{2d_i-D}{\sqrt{(D-1)(D-2)}}\phi} T\left[
F^{(i)},g\right]\; ,
\ee
where $T_{ab}\left[ F^{(i)},g\right]$, $T\left[ F^{(i)},g\right]$
are given by replacing $\ov g \longrightarrow g$ in Eqs. \rf{1.7},
\rf{1.9}. The indices of the field strengths $F^{(i)}$ are now
raised and lowered with the metric $g$.

The equations of motion \rf{1.10} for $F^{(i)}$ transform to
\be{1.23}
\frac1{\sqrt{|g|}} \left( \sqrt{|g|}\;
e^{\frac{2d_i-D}{\sqrt{(D-1)(D-2)}}\phi}\; F^{(i)\, m_i n_i \dots
q_i}\right)_{,\, q_i}=0\, ,\quad i=1,\dots ,n\, ,
\ee
whereas the Bianchi identities \rf{1.11} do not change.

It can be easily checked that Eqs. \rf{1.17}, \rf{1.18} and
\rf{1.23} are the equations of motion for the action
\be{1.24} S = \frac{1}{2\kappa_D^2} \int_M d^D x \sqrt{|g|}
\left\{ R[g] - g^{ab} \phi_{,a} \phi_{,b} - 2 U(\phi )-
\kappa^2_D\sum_{i=1}^n \frac1{d_i!} \;
e^{\frac{2d_i-D}{\sqrt{(D-1)(D-2)}}\phi}\;F^{(i)}_{m_i n_i \dots
q_i} F^{(i)\, m_i n_i \dots q_i}\right\}\; , \ee
where
\be{1.25} U(\phi ) := \frac12 e^{- B \phi} \left[\; \ov R (\phi
)e^{A \phi } - f\left( \ov R (\phi )\right) \right]\; , \quad B :=
\frac {D}{\sqrt{(D-1)(D-2)}}\;  \ee
and Eq. \rf{1.19} is used to express $\ov R$ as a function of
$\phi$ : $\ov R = \ov R( \phi )$. The scalar field $\phi$ is the
result and the carrier of the curvature nonlinearity of the
original theory\footnote{Thus, for brevity, we shall refer to the
field $\phi$ as nonlinearity scalar field. \label{nonlinear scalar
field}} \rf{1.1}. Correspondingly, Eq. \rf{1.18} has a two-fold
interpretation. It is the equation of motion for the field $\phi$
and at the same time it can be considered as constraint equation
following from the reduction of the nonlinear theory \rf{1.1} to
the linear one \rf{1.24}. Furthermore, we note that in the linear
theory \rf{1.24} the form fields are non-minimally coupled with
the nonlinearity field $\phi$. (A minimal coupling occurs only for
a model with $n=1$, $d_1=D_0$, where according to \rf{1.9} the
trace of the form field EMT vanishes.) A comparison of the action
functional  with \rf{1.21} shows that the last term in \rf{1.24}
coincides with the expression for the energy density
$-T^0_0[F,\phi,g]$ of the solitonic form field (due to
$F^{(i)}_{0n_i\ldots q_i}\equiv 0$ by the definition of
$F^{(i)}$).

Let us consider what happens if, in some  way, the scalar field
$\phi$ tends asymptotically to a constant: $\phi \to \phi_{0} $
[precisely this situation should hold when the internal space
undergoes a (freezing) stabilization]. {}From Eq. \rf{1.19} we see
that in this limit the nonlinearity in \rf{1.1} disappears: $f(\ov
R) \approx c_1(\ov R - \ov R_0) + f(\ov R_0) \equiv c_1 \ov R +
c_2$, where $c_1 := f'(\ov R_0) = \exp(A \phi_{0})$, \ $\ov R_0
\equiv \ov R (\phi_0)$,  and $-c_2/(2c_1)$ plays the role of a
cosmological constant. In the case of homogeneous and isotropic
spacetime manifolds, linear purely geometrical theories with
constant $\Lambda -$term necessarily imply an (A)dS geometry so
that the manifolds are Einstein spaces. In our model, the
additional form fields destroy this asymptotical behavior.
Instead, we obtain from Eqs. \rf{1.5} and \rf{1.4}
\be{1.25a} \ov R \longrightarrow  -\frac{D}{D-2}\frac{c_2}{c_1} -
\frac{1}{c_1}\frac{2}{D-2} \kappa^2_D \sum_{i=1}^{n}
\frac{2d_i-D}{2(d_i!)}\left(F^{(i)}\right)^2_{\ov g} \ee
and
\be{1.25b} \ov R_{ab} \longrightarrow \left[
-\frac{1}{D-2}\frac{c_2}{c_1} - \frac{1}{c_1}\frac{2}{D-2}
\kappa^2_D \sum_{i=1}^{n} \frac{d_i-1}{2(d_i!)}\left( F^{(i)}
\right)^2_{\ov g}\right]\ov g_{ab} + \frac{1}{c_1}\kappa^2_D
\sum_{i=1}^{n} \frac{2d_i}{2(d_i!)}\left( F^{(i)}_{a n_i \dots q_i}
F^{(i)\, n_i\dots q_i}_b \right)_{\ov g} \, ,\ee
where the form field product
\be{1.25c} \left(F^{(i)}\right)^2_{\ov g} := F^{(i)}_{m_i n_i\dots q_i}
F^{(i)\, m_i n_i\dots q_i}\ee
is defined with respect to the metric $\ov g$. For a
 form field, which asymptotically tends to a constant,
the scalar curvature and the Ricci tensor also approach constant
values.  But whereas $\ov R$ and $R$ are asymptotically connected
by the relation [see Eqs. \rf{1.16}, \rf{1.17} and \rf{1.25a}]
\be{1.25d} R \longrightarrow c_1^{-\frac{2}{D-2}}\, \ov R \, , \ee
the Ricci tensor $\ov R_{ab}$ will not be proportional to the
metric $\ov g_{ab}$ and, hence, the space will not be Einsteinian.
This is in obvious contrast to a nonlinear model of purely
geometrical type \cite{GMZ(PRDa)} where the stabilization will
result in an asymptotical $\mbox{(A)dS}_D$ spacetime.

In the rest of the paper we consider for simplicity a toy
model\footnote{For considerations on higher order corrections to
the gravity sector of M/string theory we refer to \cite{EKOY,r4}.}
with quadratic curvature term:
\be{1.26} f(\ov R ) = \ov R + \alpha \ov R^{\; 2} - 2\Lambda _D\;
,\ee
where the parameter $\alpha$ has dimension $\mathcal{O}(m^{-2})$.
For this model we obtain
\be{1.27} f'(\ov R )= 1 + 2\alpha \ov R = e^{A \phi} \Longleftrightarrow \ov R
= \frac{1}{2\alpha } \left( e^{A \phi } - 1\right) \ee
and
\be{1.28} U(\phi ) = \frac12 e^{-B \phi }\left[ \frac{1}{4\alpha
}\left( e^{A \phi } - 1\right)^2 + 2 \Lambda _D \right]\; . \ee
The condition\footnote{Obviously, the conformal transformation
\rf{1.12}, \rf{1.13} becomes singular when $f'(\ov R)$ vanishes.
The transformation itself can be extended from the
 $f'>0$ branch to the $f'<0$ branch with the help of an ansatz
\cite{Maeda} $\Omega = |f' (\ov R)|^{1/(D-2)}$ and a corresponding
redefinition of the nonlinearity field $\phi$: $e^{A\phi} = |f'|$.
As result, one obtains an action functional for the $f'<0$ branch
which differs from action \rf{1.24} for $f'>0$ in its total sign
and in the relative signs of the single terms as well as in the
potential $U(\phi)$. Most important, for a fixed sign of the
Einstein-Hilbert term the kinetic term of the nonlinearity field
has the correct sign, whereas the kinetic terms of additional
matter fields (in our case the form fields) have the wrong
relative sign. This leads to a set of equations of motions which
differ from \rf{1.17}, \rf{1.18}. For details we refer to
\cite{Maeda}. Unfortunately, the sign switch occurs for all
additional matter fields simultaneously and it is not controllable
for some selected fields separately. Otherwise, it could have
provided a natural mechanism for the generation of a phantom
energy component with equation of state parameter $\omega_Q<-1$
(and a corresponding super-acceleration of the observable
Universe) in the sense of \cite{phantom}.\label{f'}} $f'
> 0$ implies $1+2\alpha \ov R > 0$. In the limit $\alpha \to
0$ the nonlinearity is switched off and the linear theory is
recovered. Correspondingly, it holds $f' \to 1$ with implication
$c_1 =1$,\ $\phi_0 \to 0$ so that also $\ov R \to R$ (in
accordance with Eq. \rf{1.25d}) and $U(\phi \to 0) \to \Lambda_D$.
The corresponding region of weak nonlinearity  is defined by the
condition $\alpha \ov R=e^{A\phi}-1\ll 1$.


\section{Spontaneous compactification and dimensional reduction \label{compact}}
\setcounter{equation}{0}

The simple block-orthogonal structure of the field strength $F$
shows that there is a preferable scheme for a spontaneous
compactification of the multidimensional spacetime manifold: each
of the form fields $F^{(i)}$ is nested in its own
$d_i$-dimensional space $M_i$. Thus, the $D$-dimensional spacetime
$M$ can be endowed with the structure of a warped product manifold
\be{2.1} M \longrightarrow M = M_0 \times M_1 \times \ldots \times
M_n \ee
with metric
\be{2.2} \ov g=\ov g_{ab}(X)dX^a\otimes dX^b=\ov
g^{(0)}+\sum_{i=1}^n \ov g^{(i)}\, . \ee

The coordinates on the $(D_0=d_0+1)$ - dimensional manifold $M_0 $
(usually interpreted as our $(D_0=4)$-dimensional Universe) are
denoted by $x$ and the corresponding metric by
\be{2.3} \ov g^{(0)}=\ov g_{\mu \nu }^{(0)}(x)dx^\mu \otimes
dx^\nu\; . \ee
Let the internal factor manifolds $M_i$ be $d_i$-dimensional
warped Einstein spaces with warp factors $\exp (\ov \beta^i (x))$
and metrics
\be{2.4} \ov g^{(i)}= e^{2 \ov \beta^i (x)}
\gamma^{(i)}_{m_in_i}(y_i)dy_i^{m_i}\otimes dy_i^{n_i}\, ,\ee
i.e.,
\be{2.5} R_{m_in_i}\left[ \gamma^{(i)}\right] =\lambda
^i\gamma_{m_in_i}^{(i)},\qquad m_i,n_i=1,\ldots ,d_i\ee
and
\be{2.5a}R\left[ \gamma^{(i)}\right] =\lambda ^id_i\equiv R_i \sim
k r_i^{-2}\; , \ee
where $k = 0,\pm 1$.  The scale $r_i$
sets the characteristic size of $M_i$ (modulo the warp factor
$\exp (\ov \beta^i)$) and can be interpreted as an effective scale
factor of the compact Einstein space $M_i$ with metric $\gamma^i$
and corresponding volume\footnote{The volume is well defined for
positive curvature spaces ($k=+1$). For compact negative and zero
curvature spaces ($k=-1,0$), i.e.  compact hyperbolic spaces
(CHSs) $M_i=H^{d_i}/\Gamma_i$ and tori $T_j=R^{d_j}/\Gamma_j$, we
interpret this volume as scaled volume of the corresponding
fundamental domain ("elementary cell") $V_{d_i} \sim
r_i^{d_i}\times V_{FD(i)}$ (see, e.g., \cite{SS} and references
therein).  Here $H^{d_i}$, $R^{d_j}$ are hyperbolic and flat
universal covering spaces, and $\Gamma_i$, $\Gamma_j$ ---
appropriate discrete groups of isometries. Furthermore, we assume
for the scale factors of the metrics $\gamma^i\sim r_i
\hat{\gamma}^i$ with $\hat{\gamma}^i$ scaled in such
a way that $V_{FD(i)}\sim \mathcal{O}(1)$.
  Thus, the volume $V_{d_i}$
is mainly defined by $r_i$. In all three cases ($k=\pm 1, 0$), the
limit $r_i \to \infty $ results in an effective decompactification
of the internal space with $V_{d_i} \to \infty$. In accordance
with Eq. \rf{2.5a}, this means that the positive and negative
constant curvature spaces flatten:  $R_{d_i} \to 0$. Clearly, for
compact Ricci-flat spaces holds $R_{d_i}\equiv 0$ by definition
and without relation to the compactification scale of the torus.
\label{V_d}}
\be{2.5b} V_{d_i } \equiv \int\limits_{M_i}d^{d_i}y
\sqrt{|\gamma^{(i)}|} \sim r_i^{d_i}\quad i=1, \dots , n \, , \ee
where $V_{d_i}$ has dimension $\mathcal{O}(m^{-d_i})$.

We note that the specific metric ansatz \rf{2.2} - \rf{2.4} for
the warped product of Einstein spaces results in a scalar
curvature $\ov R$ which depends only on $x$: $\ov R[\ov g] = \ov
R(x)$. Correspondingly, the nonlinearity field $\phi$ is also a
function only of $x$: $\phi = \phi (x)$.

The conformally transformed metric \rf{1.12} reads
\be{2.6} g = \Omega^2 \ov g = \left( e^{A \phi }\right)^{2/(D-2)}
\ov g\: := g^{(0)}+\sum_{i=1}^ne^{2 \beta^i(x)}\gamma^{(i)}\;  \ee
with
\be{2.7} g^{(0)}_{\mu \nu} = \left( e^{A \phi}\right)^{2/(D-2)}
\ov g^{(0)}_{\mu \nu}\; , \ee \be{2.8} \beta^i = \ov {\beta}^i +
\frac{A}{D-2} \phi\; . \ee

For the field strengths $F^{(i)}$ we choose a generalized
Freund-Rubin ansatz \cite{FR} (see also
\cite{wiltshire,LOW,IM1a,GC1}):
\be{2.9} F^{(i)}_{m_i n_i\dots q_i} = \sqrt{2}\; \sqrt{|g^{(i)}|}\;
\varepsilon_{m_i n_i\dots q_i} f^{(i)}(x), \qquad  F^{(i)\; m_i
n_i\dots q_i} = \left(\sqrt{2} / \sqrt{|g^{(i)}|}\right)
\varepsilon^{m_i n_i\dots q_i} f^{(i)}(x)\; ,
\ee
where $g^{(i)} \equiv e^{2\beta^i} \gamma^{(i)}$ and
$\varepsilon_{m_i n_i\dots q_i}$ is the Levi-Civita symbol. We use
conventions where for Riemann spaces holds $\varepsilon_{m_i
n_i\dots q_i} = \varepsilon^{m_i n_i\dots q_i}$ and
$\varepsilon_{m_i n_i\dots q_i}\varepsilon^{m_i n_i\dots q_i}
=d_i\,!$. It can be easily seen that the ansatz \rf{2.9} satisfies
Eq. \rf{1.23} (because $\phi$ and $f$ depend only on $x$ and the
$\sqrt{|\gamma^{(i)}|}$ factors cancel). The Bianchi identities
\rf{1.11} reduce
 to the equations
\be{2.10-a} \frac{\partial\left( a^{d_i}_i(x)
f^{(i)}(x)\right)}{\partial x^{\mu}} = 0
\ee with solutions
\be{2.10}
 f^{(i)}(x) =
\frac{f_i}{a^{d_i}_i}\, , \ee
where $a_i := e^{\beta^i}$ are the scale factors of the internal
spaces $M_i$ after conformal transformation \rf{2.6} and
$f_i\equiv \const$. We choose the scale factors $a_i$
dimensionless so that the constants $f^2_i$ have dimension
$\mathcal{O}(m^{4+D'})$ and $\kappa^2_D f_i^2 \sim
\mathcal{O}(m^{2})$. With \rf{2.10} the energy density of the
solitonic form field, and correspondingly the last term in action
\rf{1.24}, reads
\be{2.11} -T^0_0[F,\phi,g]=\frac12\sum_{i=1}^n \frac1{d_i!} \;
e^{\frac{2d_i-D}{\sqrt{(D-1)(D-2)}}\phi}\;F^{(i)}_{m_i n_i \dots
q_i} F^{(i)\, m_i n_i \dots q_i} = \sum_{i=1}^n \,
e^{\frac{2d_i-D}{\sqrt{(D-1)(D-2)}}\phi}\;
\frac{f_i^2}{a_i^{2d_i}} :=\rho (\beta , \phi) \; , \ee
where for real form fields $f_i^2 \ge 0$. Again we see that for
models with  $n=1$ and $d_1=D_0$ this energy density  decouples
from the nonlinearity scalar field $\phi$: $\rho(\beta^1, \phi )
\longrightarrow \rho(\beta^1)$.

The fact that $\phi $, $\beta^i $ and $\rho$ depend only on $x$
allows us to perform a dimensional reduction of action \rf{1.24}.
Without loss of generality we set the compactification scales of
the internal spaces $M_i\quad i = 1,\ldots ,n$ at present time at
$\beta^i = 0$. This means that at present time the total volume of
the internal spaces is completely defined by the characteristic
scale factors $r_i$ (see \rf{2.5b} and footnote \rf{V_d}):
\be{2.5c}
 V_{D^{\prime }} \equiv
\prod_{i=1}^n\int\limits_{M_i}d^{d_i}y \sqrt{|\gamma^{(i)}|} \sim
\prod_{i=1}^n r_i^{d_i}\; ,
\ee
where $D^{\prime} = D-D_0 = \sum_{i=1}^n d_i$ is the number of
extra dimensions and $V_{D^{\prime }}$ has dimension
$\mathcal{O}(m^{-D^{\prime}})$. After dimensional reduction action
\rf{1.24} reads
\ba{2.12}
S&=&\frac 1{2\kappa _0^2}\int\limits_{M_0}d^{D_0}x\sqrt{|g^{(0)}|}%
\prod_{i=1}^ne^{d_i\beta ^i}\left\{ R\left[ g^{(0)}\right]
-G_{ij}g^{(0)\mu \nu }\partial _\mu \beta ^i\,\partial _\nu \beta
^j -g^{(0)\mu \nu}\partial_{\mu} \phi \partial_{\nu } \phi \right.
\nn\\
 && +\sum_{i=1}^n\left. R\left[ g^{(i)}\right] e^{-2\beta
^i}-2 U(\phi) - 2\kappa^2_D\, \rho (\beta ,\phi )\right\} \; .
\ea
where  $G_{ij}=d_i\delta _{ij}-d_id_j\ (i,j=1,\ldots ,n)$ is the
midisuperspace metric \cite{IMZ, RZ} and
\be{2.13} \kappa^2_0 := \frac{\kappa^2_D}{V_{D^{\prime }}} \ee
denotes the $D_0-$dimensional (4-dimensional) gravitational
constant. If we take the electroweak scale $M_{EW}$ and the Planck
scale $M_{Pl(4)}$ as fundamental ones for $D-$dimensional (see Eq.
\rf{1.1a}) and 4-dimensional spacetimes ($\kappa^2_0 = 8\pi /
M^2_{Pl(4)}$) respectively, then we reproduce Eqs. \rf{0.1} and
\rf{0.2}.

Action \rf{2.12} is written in the Brans - Dicke frame. Conformal
transformation to the Einstein frame \cite{GZ1,GZ}
\be{2.14} \td g_{\mu \nu }^{(0)}= {\left( \prod_{i=1}^ne^{d_i\beta
^i}\right) }^{\frac 2{D_0-2}}g_{\mu \nu }^{(0)} \ee
yields
\be{2.15} S=\frac 1{2\kappa
_0^2}\int\limits_{M_0}d^{D_0}x\sqrt{|\td g^{(0)}|}\left\{ R\left[
\td g^{(0)}\right] -\bar G_{ij}\td g^{(0)\mu \nu }\partial _\mu
\beta ^i\,\partial _\nu \beta ^j- \td g^{(0) \mu \nu}
\partial_{\mu}\phi \partial_{\nu} \phi -2U_{eff} (\beta ,\phi )
\right\} \; . \ee
The tensor components of the midisuperspace metric (target space
metric on $\RR _T^n$ ) $\bar G_{ij}\ (i,j=1,\ldots ,n)$ , its
inverse metric $\bar G^{ij}$ and the effective potential are
correspondingly
\be{2.16} \bar G_{ij}=d_i\delta _{ij}+\frac 1{D_0-2}d_id_j\;
,\quad \bar G^{ij}=\frac{\delta ^{ij}}{d_i}+\frac 1{2-D} \ee
and
\be{2.17} U_{eff}(\beta ,\phi ) ={\left( \prod_{i=1}^n e^{d_i\beta
^i}\right) }^{-\frac 2{D_0-2}} \left[ -\frac
12\sum_{i=1}^nR_ie^{-2\beta ^i}+ U (\phi ) + \kappa^2_D\, \rho
(\beta ,\phi ) \right] \; , \ee
where $U (\phi )$ and $\rho (\beta ,\phi )$ are defined
 by Eqs. \rf{1.28} and \rf{2.11}.

A stable compactification of the internal spaces is ensured when
the scale factors of the internal spaces $\beta^i$ are frozen at
one of the minima of the effective potential $U_{eff}$. The value
of the effective potential at the minimum plays the role of the
effective $D_0-$dimensional cosmological constant: $\left.
U_{eff}\right|_{min} \equiv \Lambda_{eff}$. Assuming for the
frozen scale factors at present time $\beta^i=0$, we obtain the
non-zero components of the asymptotic Ricci tensor \rf{1.25b} as
\ba{2.18curv}\ov{ R}_{\mu \nu}&\longrightarrow &\vartheta \; \ov{ g}_{\mu
\nu}\; ,\label{2.18curv-1}
\\\ov{ R}_{m_i n_i}&\longrightarrow &\left(\vartheta
+\frac{2}{c_1}\kappa_D^2 f_i^2 \Omega_0^{2d_i}\right)\ov{ g}_{m_i
n_i}\; ,\label{2.18curv-2}
\ea
where
\be{2.19curv}
\vartheta :=
-\frac{1}{D-2}\frac{c_2}{c_1}
-\frac{1}{c_1}\frac{2}{D-2}\kappa_D^2\sum_{j=1}^n(d_j-1)f_j^2\Omega_0^{2d_j}
\ee
and $\Omega_0=\left(e^{A\phi_0}\right)^{1/(D-2)}$. Thus, the
asymptotic multidimensional spacetime is built up from
Einstein-space blocks, but is itself a non-Einsteinian space due
to the additional term in \rf{2.18curv-2}.


\section{Stabilization of the internal space \label{stability}}
\setcounter{equation}{0}

Without loss of generality\footnote{The difference between a
general model with $n>1$ internal spaces and the particular one
with $n=1$ consists in an additional diagonalization of the
geometrical moduli excitations. \label{n=1}}, we consider in the
present section a model with only one $d_1$-dimensional internal
space. The corresponding action \rf{2.15} reads
\be{3.1} S=\frac 1{2\kappa
_0^2}\int\limits_{M_0}d^{D_0}x\sqrt{|\td g^{(0)}|}\left\{ R\left[
\td g^{(0)}\right] - \td g^{(0) \mu \nu}
\partial_{\mu}\varphi \partial_{\nu} \varphi - \td g^{(0) \mu \nu}
\partial_{\mu}\phi \partial_{\nu} \phi -2U_{eff} (\varphi ,\phi )
\right\} \, , \ee
where
\be{3.2} \varphi := -\sqrt{\frac{d_1(D-2)}{D_0-2}}\beta^1 \ee
and
\be{3.3} U_{eff}(\varphi ,\phi ) = e^{2\varphi
\sqrt{\frac{d_1}{(D-2)(D_0-2)}}} \left[ -\frac12 R_1e^{2\varphi
\sqrt{\frac{D_0-2}{d_1(D-2)}}}+ U (\phi ) + \kappa^2_D\, \rho
(\varphi ,\phi ) \right] \; . \ee
The potential $U(\phi )$ of the  nonlinearity scalar field is
given by Eq. \rf{1.28} and the energy density \rf{2.11} of the
solitonic form field  reads
\be{3.4} \kappa^2_D\, \rho (\varphi ,\phi )\, =\,
\kappa^2_D\,f^2_1\, e^{\frac{2d_1-D} {\sqrt{(D-1)(D-2)}}\phi }\,
e^{2\varphi \sqrt{\frac{d_1(D_0-2)}{D-2}}}.
\ee
For brevity of the notation, we introduce
\ba{3.4-1}
a:=2\sqrt{\frac{D_0-2}{d_1(D-2)}}\; , &&
b:=2\sqrt{\frac{d_1}{(D-2)(D_0-2)}}\; ,\nn
\\
 c:= \frac{2d_1-D}{\sqrt{(D-1)(D-2)}}\; ,&& h:=\kappa^2_D\,f^2_1
\ea
so that the effective potential reads
\be{3.4-2}
U_{eff}=e^{b\varphi}\left[-\frac 12 R_1 e^{a\varphi}+U(\phi)+h
e^{c\phi}e^{ad_1 \varphi} \right]\; .
\ee
{}From \rf{3.4-1} we see that a real-valued form field $f_1$
implies a non-negative $h=\kappa^2_D\,f^2_1\ge 0$. For the rest of
the paper, we continue to work with dimensionless scalar fields
$\varphi ,\phi $ instead of passing to canonical ones (modulo
$8\pi $): $\tilde{\varphi} = \varphi \, M_{Pl(4)}, \, \tilde{\phi}
= \phi \, M_{Pl(4)} $ and $\tilde{U}_{eff} = M_{Pl(4)}^2 U_{eff}$.
The restoration of the correct dimensionality is obvious.

In order to ensure a stabilization and asymptotical freezing of
the internal space $M_1$,  the effective potential should have a
minimum with respect to the scalar field $\varphi$
\be{3.4-3}
\left. \partial_{\varphi} U_{eff}\right|_{extr}=0\; ,
\ee
so that for a minimum position at $\varphi_0=0$
(which corresponds to a compactification
scale $\beta^1=0$ at present time) it should hold
\be{3.4-4}
\frac{a+b}{2} R_1 =b U(\phi)+(ad_1+b)h e^{c\phi}.
\ee
This formula shows that the potential $U_{eff}(\varphi ,\phi)$
must also have a minimum with respect to $\phi$, because without
stabilization of $\phi$ the right hand side remains a dynamical
function whereas the left hand side is a constant. This second
extremum condition
\be{3.4-5}
\left. \partial_{\phi} U_{eff}\right|_{extr}=0
\ee
yields
\be{3.4-6}
\left.\left[\partial_{\phi}U+hc e^{c\phi}\right]\right|_{extr}=0.
\ee
Additionally, the eigenvalues of the mass matrix of the coupled
$(\varphi,\phi)-$field system, i.e. the Hessian of the effective
potential at the minimum position,
\be{3.4-7}
J:=\left.\bmatrix{ccc}
\partial^2_{\varphi\varphi}U_{eff}
&&\partial^2_{\varphi\phi}U_{eff}\\ \\
\partial^2_{\phi\varphi}U_{eff}
&&\partial^2_{\phi\phi}U_{eff} \ematrix\right|_{extr}
\ee
should be positive definite
\be{3.4-8}
m^2_{1,2}=\frac12\left[Tr(J)\pm\sqrt{Tr^2(J)-4\det(J)}\right]> 0
\; .
\ee
According to the Sylvester criterion this is
 equivalent to the  conditions
\be{3.4-9}
J_{11}>0, \qquad J_{22}>0, \qquad \det(J)>0.
\ee

{}From \rf{3.4-7} we see that in the special case of
$\left.\partial^2_{\varphi\phi}U_{eff}\right|_{extr}=0$ the
Hessian is diagonal and the excitation modes of the fields
decouple. The eigenvalues of $J$ coincide in this case with the
masses squared of the scale factor excitations (gravitational
excitons \cite{GZ1}) $m_1^2=m^2_{\varphi}$ and the excitations of
the nonlinearity field $m^2_2=m^2_{\phi}$.

Let us now analyze the stability conditions \rf{3.4-4}, \rf{3.4-6}
and \rf{3.4-9} explicitly. For this purpose we introduce the
auxiliary notations
\be{3.4-10}
\phi_0:=\left.\phi\right|_{extr},\qquad X:=e^{A\phi_0}\ge 0,
\qquad q:=8\alpha \Lambda_D
\ee
and rewrite the potentials $U$, $U_{eff}$ and the derivatives of
$U_{eff}$ at a possible minimum position $(\varphi_0=0,\phi_0)$ as
\ba{3.4-11}
U_0\equiv\left.U\right|_{extr}&=&\frac{1}{8\alpha}X^{-\frac{D}{D-2}}\left[\left(X-1\right)^2+q\right]\;
,\label{3.4-11-1}\\ \left.U_{eff}\right|_{extr}&=&-\frac 12
R_1+U_0(X)+h X^{\frac{2d_1-D}{D-2}}\; ,\label{3.4-11-2}
\\
\left.\partial_{\varphi}U_{eff}\right|_{extr}&=&-\frac{a+b}{2}R_1+b
U_0(X) + (d_1 a+b)hX^{\frac{2d_1-D}{D-2}}=0\; ,\label{3.4-11-3}
\\
\left.\partial_{\phi}U_{eff}\right|_{extr}&=&\frac{1}{8\alpha}X^{-\frac{D}{D-2}}
\left[(2A-B)X^2-2(A-B)X-(q+1)B\right]+hcX^{\frac{2d_1-D}{D-2}}=0\;
,\label{3.4-11-4}\\
\left.\partial^2_{\varphi\varphi}U_{eff}\right|_{extr}
&=&-\frac{a^2-b^2}{2}R_1-b^2U_0(X)+\left[(d_1a)^2-b^2\right]hX^{\frac{2d_1-D}{D-2}}\;
,\label{3.4-11-5}\\
\left.\partial^2_{\varphi\phi}U_{eff}\right|_{extr}
&=&cd_1ahX^{\frac{2d_1-D}{D-2}}\; ,\label{3.4-11-6}\\
\left.\partial^2_{\phi\phi}U_{eff}\right|_{extr} &=&
\frac{1}{8\alpha}X^{-\frac{D}{D-2}}\left[(2A-B)^2X^2-2(A-B)^2X+(q+1)B^2\right]+c^2hX^{\frac{2d_1-D}{D-2}}\;
.\label{3.4-11-7}
\ea
(The constants $A$, $B$ are defined in Eqs. \rf{1.19} and
\rf{1.25}, respectively.) We see that, for fixed dimensions $D_0$
and $d_1$, the two equations \rf{3.4-11-3}, \rf{3.4-11-4} describe
a $3-$dimensional algebraic variety ${\cal V}\subset {\cal M}$ in
the $5-$dimensional parameter (moduli) space\footnote{The
compactification scale (modulus) $r_1$ of the internal space $M_1$
enters  ${\cal V}\subset{\cal M}$ via curvature scalar $R_1$ (see
Eq. \rf{2.5a}).} ${\cal M}=\RR^3\times \RR^2_+\ni
(\alpha,\Lambda_D,R_1,h,X)$. On the variety, inequalities
\rf{3.4-9} of the Sylvester criterion define then the parameter
region $ \Upsilon\subset {\cal V}$ of stable compactifications. A
natural strategy for extracting detailed information about the
location of this stability region would consist in solving
 \rf{3.4-11-4} for $X$ with subsequent
back-substitution of the found roots into the inequalities
\rf{3.4-9} and the equation \rf{3.4-11-3}.
In the following consideration we restrict our
attention to the three simplest nontrivial cases which are easy to
handle analytically.

\vspace{0.5cm}

\subsection{Zero effective cosmological constant:
$\Lambda_{eff} \equiv 0$ \label{zero}}

By definition,  we have $\Lambda_{eff} \equiv
\left.U_{eff}\right|_{extr}$  so that in the particular case
$\Lambda_{eff} \equiv 0$ Eq. \rf{3.4-11-2} yields the additional
constraint
\be{3.4-12}
\left.U_{eff}\right|_{extr}=-\frac 12 R_1+U_0(X)+h
X^{\frac{2d_1-D}{D-2}}=0 \; .
\ee
Combining this constraint with \rf{3.4-11-3} we obtain the
relation
\be{3.4-13}
R_1=2d_1hX^{\frac{2d_1-D}{D-2}}=\frac{2d_1}{d_1-1}U_0(X)
\ee
which can be used to eliminate the $hX^{\frac{2d_1-D}{D-2}}$ term
from \rf{3.4-11-4}. As result we arrive at a simple quadratic
equation in $X$ with physically sensible solutions
\be{3.4-14}
 e^{A\phi_0} \equiv X= \left\{\begin{array}{rcl}
\frac{-1+\sqrt{1+(d_1-2)d_1(1+q)}}{d_1-2} \, , \quad d_1>2\; ,\\
1+q\, ,\qquad d_1=2\; .\\
\end{array}\right.
\ee
With the help of Eqs. \rf{3.4-13}, \rf{3.4-14} and repeated use of
a  substitution-elimination technique, the potential $U_0$ and the
second derivatives \rf{3.4-11-5} - \rf{3.4-11-7} of the effective
potential can be rewritten in the simpler form
\be{3.4-14b}
U_0(X)=\frac{d_1-1}{4\alpha d_1}X^{-\frac{2}{D-2}}(X-1)
\ee
and
\ba{3.4-15}
 J_{11} &\equiv & \left.
\partial^2_{\varphi \varphi } U_{eff} \right|_{extr} = a^2d_1 U_0(X)\, , \label{3.4-15-1}
\\
J_{22} &\equiv & \left. \partial^2_{\phi \phi } U_{eff}
\right|_{extr} = \frac{B^2}{4\alpha D^2} X^{-\frac{2}{D-2}}
\left[E X +4(D_0-1)\right]\, ,\label{3.4-15-2}
\\ J_{12} &\equiv & \left.
\partial^2_{\varphi \phi } U_{eff} \right|_{extr} =
\frac{cd_1a}{d_1-1} U_0(X)\, ,\label{3.4-15-3}
\ea
where $E\equiv (D-4)^2+ 4(d_1-2)
>0$ for $d_1 \ge 2$. For the determinant of the Hessian \rf{3.4-7}
we get
\be{3.4-16}
\det(J) = \frac{1}{\alpha (d_1-1) }\frac{D_0-2}{D-1}\,
U_0(X)X^{-\frac{2}{D-2}}\left[(d_1-2)X +1\right]\, .
\ee

With the equations \rf{3.4-13} - \rf{3.4-16} at hand, we are well
prepared to explicitly describe the location of the stability
region $\Upsilon$ in the parameter space ${\cal M}$. Let us start
with relation \rf{3.4-13}. {}From the non-negativity conditions
$h\ge 0$ and $e^{A\phi_0}\equiv X\ge 0$ we immediately conclude
that for stable spaces $M_1$ it should hold $R_1 \ge 0$ and
$U_0(X)\ge 0$. Furthermore, we see from the latter condition and
the Sylvester criterion $J_{22}>0$, $\det (J)>0$ [applied to
\rf{3.4-15-2} and \rf{3.4-16}] that for internal spaces of
dimension $d_1\ge 2$ the parameter $\alpha$ is restricted to
positive values\footnote{Obviously, a negative $\alpha$ would
yield a maximum of the effective potential $U_{eff}$ instead of a
minimum and our model would become unstable with respect to the
conformal excitations of the internal space. The condition
$\alpha>0$ is also required in other $R^2$ models \cite{AEL} to
ensure tachyon-free configurations. \label{tachyon}} $\alpha>0$
(the limiting case $\alpha \to 0$ we discuss below). Finally, we
note that Eq. \rf{3.4-14b} together with $U_0(X)\ge 0$ and
$\alpha>0$ implies $X\ge 1$ and, hence, we find for $d_1>2$ and
$d_1=2$ from the roots \rf{3.4-14}: $q\equiv 8\alpha \Lambda_D\ge
0$ and also $\Lambda_D\ge 0$.

Summarizing the obtained restrictions, we can describe the part $
\Theta $ of the parameter space ${\cal M }$ where the stability
region $\Upsilon$ of the variety ${\cal V}$ is located:
\be{3.4-17}
\Upsilon\subset {\cal V}\cap \Theta \subset \Theta =(\alpha \ge
0,\Lambda_D\ge 0,R_1\ge 0,h\ge 0 ,X\ge 1)\subset {\cal M}\; .
\ee
It remains to clarify what happens in the various limiting cases
when the parameters reach the boundary  $\partial \Theta $.
\begin{enumerate}
\item[(L.1.1)] $q\to +0$: According to \rf{3.4-14}, this limit
implies $X\to 1$, $\phi_0\to 0$. Because of $q=8\alpha \Lambda_D$
we have to distinguish the two cases $\alpha \to 0$ and
$\Lambda_D\to 0$. In these limits we obtain $U_0(X)\to \Lambda_D$
and $U_0(X)\to 0$, respectively. \label{q0}
\item[(L.1.2)]$\alpha\to +0, \Lambda_D\neq 0$: The case $\alpha \to 0$
describes the transition to a linear model. Here we have $U(\phi
)\to \Lambda_D,\, R_1 \to [2d_1/(d_1-1)]\Lambda_D$ and $\Lambda_D
\to (d_1-1)h$. In this limit, the mass of the $\phi$-field
excitations tends to infinity $m_2^2 \to m_{\phi}^2 \to J_{22} \to
\infty$ and the field itself becomes frozen at the position
$\phi_0 \to 0$. The stabilization of the internal space occurs for
$R_1, h, \Lambda_D >0$ with the gravexciton masses $m_1^2 \to
m_{\varphi}^2 \to J_{11} = 4[(D_0-2)/(D-2)]\Lambda_D$. This is in
accordance with the results of Ref. \cite{GZ1}, where a linear
model with monopole terms was considered.
\item[(L.1.3)]$\Lambda_D\to 0, \alpha\neq 0$:
Due to \rf{3.4-14} and (L.1.1) this limit implies
$X\to 1$, $\phi_0\to 0$ and $U_0(X)\to 0$ so that according to
\rf{3.4-15-3} the excitation masses $m_1$, $m_2$ decouple
$(J_{12}\to 0)$  and the gravexciton mass
vanishes $m_1^2\to m^2_{\varphi}\to 0$.
Hence, the limit $\Lambda_D\to 0$ is connected
with a destabilization of the internal space $M_1$.
The mass of the nonlinearity field excitations $m_{\phi}$
remains finite $m_2^2\to m_{\phi}^2\sim 1/\alpha$ for $\alpha > 0$.
\item[(L.1.4)] $h\to +0, \alpha\neq 0$: {}From \rf{3.4-13} - \rf{3.4-14b}
we see that this limit of a vanishing form field is connected with
 $R_1$,$ U_0(X)$,$ \Lambda_D\to +0$. Thus the excitations of the
nonlinearity field $\phi$ decouple from  gravexcitons ($J_{12}\to
0$). Simultaneously, because of $R_1\to +0 \ \Longrightarrow
r_1\to \infty $, the internal space $M_1$ undergoes a
decompactification  and due to $U_0(X)\to 0$ the effective
potential $U_{eff}$ becomes flat in the $\varphi -$direction
($J_{11}\to 0 \Longrightarrow m_{\varphi}\to 0$). This means that
the internal space destabilizes, whereas $U_{eff}$ remains well
behaved with respect to $\phi $. These results completely confirm
the conclusions of paper \cite{GMZ(PRDa)} for a  nonlinear
gravitational model without form fields where a stabilization is
only possible for $\Lambda_{eff}<0$.
\end{enumerate}
Finally, we note that for a model with $d_1=D_0$ (and, hence, a
 vanishing trace of the form field EMT)  the
excitations of nonlinearity field $\phi$ decouple from the
gravexcitons:  $J_{12}=0$ because of $c=0$ in \rf{3.4-1},
\rf{3.4-15-3}.

\subsection{Traceless EMT of the form field:
$d_1=D_0$ \label{traceless}}

The easy handling of nonlinear models with a traceless form-field
EMT as well as of models  with a two-dimensional internal space
$M_1$ is connected with the structure of Eq. \rf{3.4-11-4}. For
$X>0$, $\alpha \neq 0$ we obtain from Eqs. \rf{3.4-1},
\rf{3.4-11-4} and the definitions of $A$ and $B$:
\be{3.5}
\frac{1}{8\alpha}\left[(D-4)X^2+4X-(q+1)D\right]+(2d_1-D)hX^{2d_1/(D-2)}=0.
\ee
This algebraic equation reduces to a simple quadratic equation in
$X$ either when the last term vanishes due to $2d_1-D\equiv
d_1-D_0=0$ (the case of a traceless form field EMT) or when its
degree $l(d_1):=2d_1/(D-2)$ equals $0,1$ or $2$. For $D_0=4$ we
have $l(d_1=0)=0$, $l(d_1=2)=1$ and $l(d_1\to \infty)\to 2$ so
that as only sensible model remains $D_0=4,d_1=2$. It will be the
subject of subsection \ref{two}.

For $d_1=D_0=4$ we find as physically sensible solution of Eq.
\rf{3.5}
\be{3.6}
X=\frac 12 \left(\sqrt{9+8q}-1\right) .
\ee
We use this solution as well as the extremum condition
\rf{3.4-11-3} to rewrite Eqs. \rf{3.4-11-1} - \rf{3.4-11-3},
\rf{3.4-11-5} - \rf{3.4-11-7} in the simpler form
\ba{3.7}
U_0(X)&=&\frac{3}{16\alpha}X^{-1/3}(X-1)\; ,\label{3.7-1}
\\ \Lambda_{eff}(X)&=&\frac 13U_0(X)-h\; ,\label{3.7-2}
\\R_1&=&4\left[\frac 13 U_0(X)+h\right]\; ,\label{3.7-3}
\\
J_{11}=\partial^2_{\varphi\varphi}\left.U_{eff}\right|_{extr}&=&\frac
23 \left[9h-U_0(X)\right]\; ,\label{3.7-4}
\\
J_{22}=\partial^2_{\phi\phi}\left.U_{eff}\right|_{extr}&=&\frac{1}{14\alpha}X^{-1/3}(2X+1)\;
,\label{3.7-5}
\\ J_{12}=\partial^2_{\varphi\phi}\left.U_{eff}\right|_{extr}&=&0\; .\label{3.7-6}
\ea
Obviously, there is no mixing of the excitations of the
nonlinearity field $\phi$ with gravexcitons $(J_{12}=0)$ in this
case: $m^2_{\varphi}=J_{11}$, $m^2_{\phi}=J_{22}$. Further, we
read off from $J_{22}>0$, $X\ge 0$ that stable internal spaces are
again only possible for $\alpha
>0$ and from \rf{3.6} and $X\ge 0$ that $q$ is restricted to the
half-line $q\ge -1$. Additional information can be extracted by
combining the condition $J_{11}>0$ with relations \rf{3.7-2},
\rf{3.7-3}, what gives \be{3.8} 16h>R_1>16U_0(X)/9>8\Lambda_{eff}.
\ee For the realistic case of a positive effective cosmological
constant we find according to \rf{3.7-2}, \rf{3.8} the conditions
\be{3.9} \Lambda_{eff}>0: \qquad h>R_1/16>U_0(X)/9>h/3>0, \ee and,
hence, from \rf{3.6}, \rf{3.7-1} also the implication $U_0(X)>0 \
\Longrightarrow \ X>1 \ \Longrightarrow q>0$. We therefore
conclude that such configurations are only possible for internal
spaces with positive scalar curvature $R_1>0$ and positive bulk
cosmological constant $\Lambda_D>0$.

Let us briefly comment on some limiting cases.
\begin{enumerate}
\item[(L.2.1)] $h\to +0$: In this case we recover the result of
\cite{GMZ(PRDa)} that stable configurations are only possible for
$R_1,\Lambda_D,\Lambda_{eff}<0$ \ [see the inequality chain
\rf{3.8}].
\item[(L.2.2)] $\alpha\to +0, \Lambda_D \neq 0$:
For this transition to the linear model with freezing of the
nonlinearity field at $\phi_0\to 0$ and diverging excitation mass
$m^2_{\phi}\to \infty$, the stability sector $\Theta \subset {\cal
M}$ can be read off from \rf{3.8} via substitution $U_0(X)\to
\Lambda_D$.
\item[(L.2.3)] $\Lambda_D \to 0, \alpha \neq 0$:
The limit is connected with $q\to 0$, $X\to 1$, $\phi_0\to 0$,
$U_0(X)\to 0$ and we have to distinguish two special cases. For a
nonvanishing form field strength $h\neq 0$ according to
\rf{3.7-4}, \rf{3.7-5} inequalities $J_{11},J_{22}>0$ hold so that
both excitation masses remain finite. For vanishing field strength
$h\to +0$ we obtain $R_1\to 0$,
 $J_{11}\to 0$, $m_1^2=m_{\varphi}^2\to 0$ and the internal space $M_1$
undergoes a destabilization/decompactification with $r_1\to
\infty$.
\end{enumerate}

\subsection{Two-dimensional internal spaces:
$d_1=2$ \label{two}}

According to Eqs. \rf{3.4-11-4}, \rf{3.5}, the extremum condition
$\left.\partial_{\phi}U_{eff}\right|_{extr}=0$ for models with
two-dimensional internal space $M_1$ and $D_0=4$ can be reduced to
a quadratic equation and, hence, allows for an easy analytical
handling of the models. Introducing the notation
\be{5.1}
z:=4\alpha h
\ee
Eq. \rf{3.5} reads
\be{5.2}
X^2-2(z-1)X-3(q+1)=0
\ee
and has solutions
\be{5.3}
X_{1,2}=z-1\pm \sqrt{(z-1)^2+3(q+1)}.
\ee
Furthermore, Eq. \rf{3.5} can be used to simplify the elements of
the Hessian $J$. Setting $D_0=4$ and $d_1=2$ everywhere in
\rf{3.4-1}, \rf{3.4-11-5} - \rf{3.4-11-7} and eliminating $q$ with
the help of \rf{5.2} we obtain\footnote{The curvature term in
$J_{11}$ of Eq. \rf{3.4-11-5} cancels because of $a=b=1$ for
$D_0=4$, $d_1=2$.}
\ba{5.4}
 J_{11} &\equiv & \left.
\partial^2_{\varphi \varphi } U_{eff} \right|_{extr} = \frac{1}{6\alpha}X^{-1/2}\left(5z+1-X\right)\, , \label{5.4-1}
\\
J_{22} &\equiv & \left. \partial^2_{\phi \phi } U_{eff}
\right|_{extr} = \frac{1}{10\alpha } X^{-1/2} \left( X
-z+1\right)\, ,\label{5.4-2}
\\ J_{12} &\equiv & \left.
\partial^2_{\varphi \phi } U_{eff} \right|_{extr} =-
\frac{1}{2\sqrt 5 \alpha} X^{-1/2}z\, \label{5.4-3}
\ea
as well as
\be{5.5}
\det(J)=-\frac{1}{60\alpha^2 X}\left(X^2-6zX+8z^2-4z-1\right)\; .
\ee
 It is now easy to describe the part $\Theta $ of the parameter space ${\cal M} $ where the stability
 region $\Upsilon \subset {\cal V}$ is located.
We start by substituting the solutions \rf{5.3} into $J_{22}$.
Taking into account that $X\ge 0$, we get from the condition
$J_{22}>0$ for $X_1$, $X_2$:
\be{5.6}
\pm \sqrt{(z-1)^2+3(q+1)}/\alpha >0\; .
\ee
Thus, the roots $X_1$ and $X_2$   correspond to $\alpha>0$ and
$\alpha <0$, respectively. Because of $z=4\alpha h$, $h\ge 0$ this
leads to negative values for $X_2$ so that this root is
 unphysical and stable configurations are restricted to $X=X_1(z,q)$ and $\alpha
>0$. The limiting case $\alpha \to +0$ will be considered separately below.

\begin{figure}[hbt]
\centerline{\epsfxsize=7cm \epsfbox{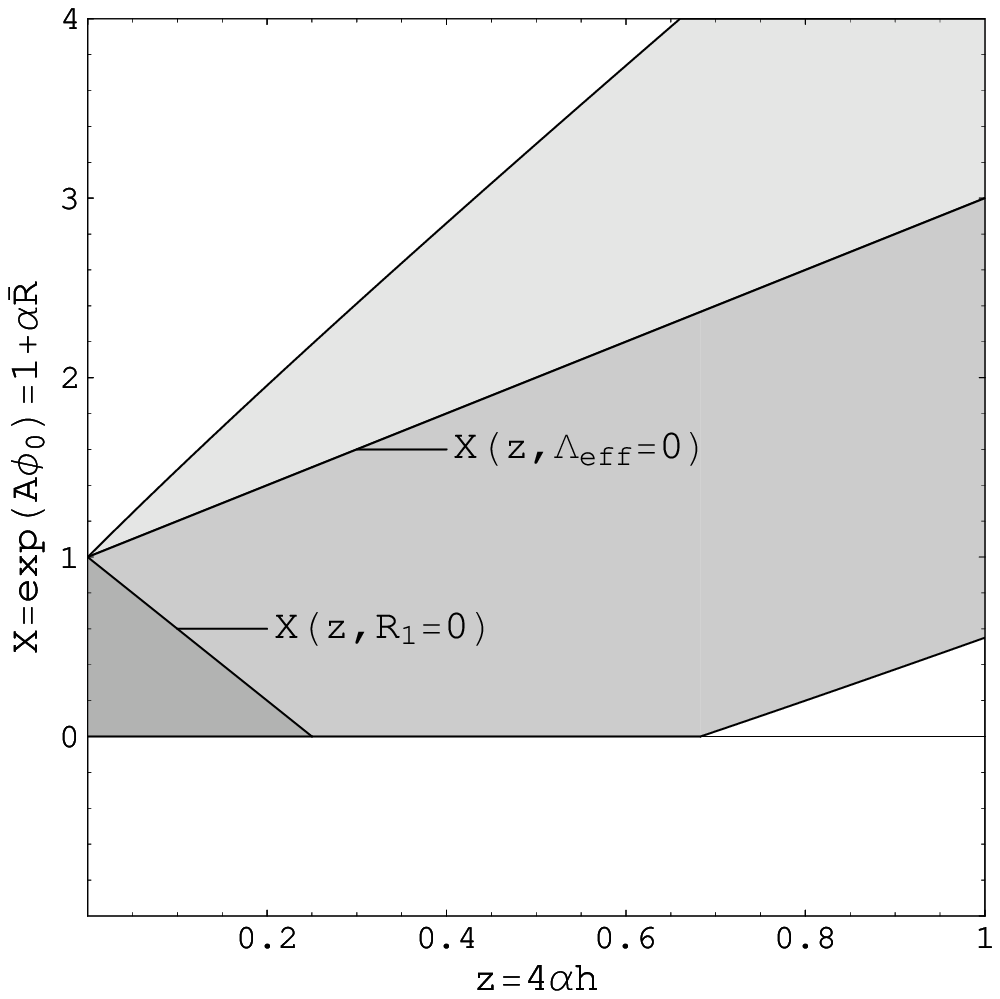}}
\caption{Projection $\Theta_{(z,X)}$ of the stability region
$\Upsilon \subset {\cal V}\subset {\cal M}$  on the $(z,X)-$plane
(shaded areas). The two lines $\Lambda_{eff}=0$ and $R_1=0$ (given
in Eqs. \rf{5.11}) separate the stable regions with:
$\left[X>X(z,\Lambda_{eff}=0):\right.$ \ $\left.(\Lambda_{eff}>0,
\right.$ $\left. R_1>0)\right]$,
$\left[X(z,\Lambda_{eff}=0)>X>X(z,R_1=0): \right.$ \ $\left.
(\Lambda_{eff}<0,\right.$ $\left. R_1>0)\right] $ and
$\left[X<X(z,R_1=0): \right.$ \ $\left. (\Lambda_{eff}<0,\right.$
$\left. R_1<0)\right] $.\label{fig1}}
\end{figure}

Furthermore, we see from the structure of Eqs. \rf{5.4-1} -
\rf{5.5} that the Sylvester criterion selects a region
$\Theta_{(z,X)}$ from the $(z,X)-$plane which can be interpreted
as the projection of the
 stability region $\Upsilon$ on this plane.
Explicitly we have
\ba{5.7}
J_{11}>0 &\Longrightarrow & X<5z+1\; , \label{5.7-1}
\\ J_{22}>0 &\Longrightarrow & X>z-1\; , \label{5.7-2}
\\ \det(J)>0 &\Longrightarrow & \left\{\begin{array}{l}X<3z+\sqrt{z^2+4z+1}\; ,\\
 X>3z-\sqrt{z^2+4z+1}\; ,  \end{array}\right. \label{5.7-3}
\ea
where the inequalities \rf{5.7-3} are easily derived from \rf{5.5}
by calculating the critical values $X_c(z)$ for which \\
$\det(J[z,X_c(z)])=0$. The intersection $\Theta_{(z,X)}$ of the
sectors defined by the conditions \rf{5.7-1} - \rf{5.7-3} and $X>
0$, $z\ge 0$ is shown on Fig. \ref{fig1}.

\begin{figure}[hbt]
\centerline{\epsfxsize=7cm \epsfbox{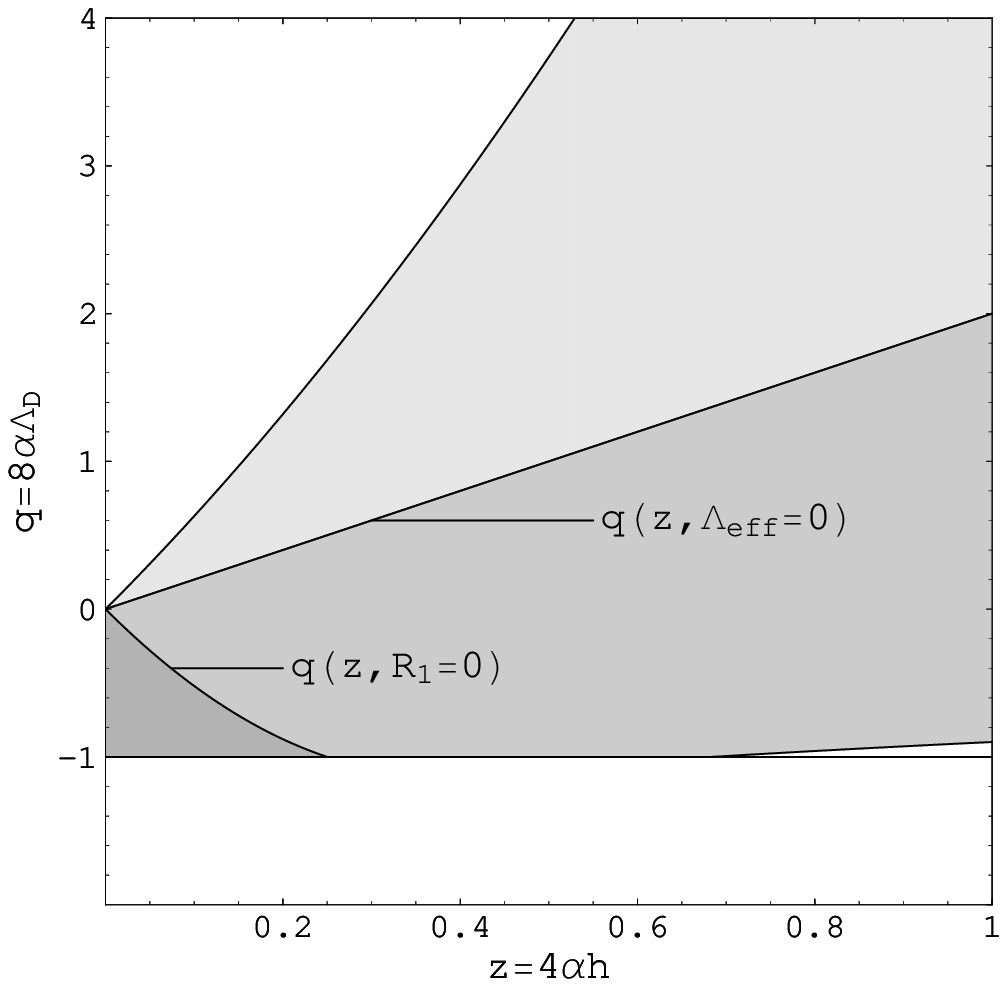}}
\caption{\label{fig2} Projection $\Theta_{(z,q)}$ of the stability
region $\Upsilon \subset {\cal V}\subset {\cal M}$  on the
$(z,q)-$plane (shaded areas). The two lines $\Lambda_{eff}=0$ and
$R_1=0$ (given in Eqs. \rf{5.11a}) separate the stable regions
with: $\left[q>q(z,\Lambda_{eff}=0): \right.$ \ $\left.
(\Lambda_{eff}>0,\right.$ $\left. R_1>0)\right]$,
$\left[q(z,\Lambda_{eff}=0)>q>q(z,R_1=0): \right.$ \ $\left.
(\Lambda_{eff}<0,\right.$ $\left. R_1>0)\right] $ and
$\left[q<q(z,R_1=0): \right.$ \ $\left. (\Lambda_{eff}<0,\right.$
$\left. R_1<0)\right] $. }
\end{figure}

In order to obtain information about the values of $q=8\alpha
\Lambda_D$ (and $\Lambda_D$) which allow for a stable internal
space $M_1$ it proves convenient to map the region
$\Theta_{(z,X)}$  via quadratic equation \rf{5.2} or its solution
$X_1(z,q)$ from the $(z,X)-$plane on an equivalent region
$\Theta_{(z,q)}$ of the $(z,q)-$plane. For this purpose it is
sufficient to transform the inequalities \rf{5.7-1} - \rf{5.7-3}
and $X\ge 0$, $z\ge 0$ for $X$ and $z$ into an equivalent
inequality set for $q$ and $z$. Let us demonstrate the mapping,
e.g., for inequality \rf{5.7-1}. Substituting
$X=X_1(z,q)=z-1+\sqrt{(z-1)^2+3(q+1)}$ into the equation for the
critical line $X=X_c(z)=5z+1$ and solving for $q$ we obtain as
image of this line $X_c(z)$ a corresponding critical curve
$q_c(z)=z(5z+6)$ on the $(z,q)-$plane. (The same curve can be
obtained by substitution of $X_c(z)$ into the quadratic equation
\rf{5.2}.) With the help of two test points
$P_1=\left(z_1,q_1>q_c(z_1)\right)$,
$P_2=\left(z_2,q_2<q_c(z_2)\right)$ above and below the critical
curve $q_c(z)$, e.g. $P_1=(1,26)$, $P_2=(2,0)$, it is then easily
seen that $X_1(z,q)<5z+1$ maps into $q<z(5z+6)$. Applying the same
technique to \rf{5.7-2}, \rf{5.7-3} we obtain \ba{5.8} J_{11}>0
&\Longrightarrow & q<z(5z+6)\; , \label{5.8-1}
\\ J_{22}>0 &\Longrightarrow & q>-1-\frac 13 (z-1)^2\; , \label{5.8-2}
\\ \det(J)>0 &\Longrightarrow &
\left\{\begin{array}{l}q<-1+\left[4z^2+10z+1+2(2z+1)\sqrt{z^2+4z+1}\right]/3\;
 ,\\
 q>-1+\left[4z^2+10z+1-2(2z+1)\sqrt{z^2+4z+1}\right]/3\; .  \end{array}\right. \label{5.8-3}
\ea
Additionally we find from $X\ge 0$
\be{5.9}
\begin{array}{lcl}q\ge -1-\frac 13 (z-1)^2 & \mbox{for} & z\ge 1\; ,\\
q\ge -1 &\mbox{for} & 0\le z \le 1\; . \end{array}
\ee
The resulting intersection region $\Theta_{(z,q)}$ of Eqs.
\rf{5.8-1} - \rf{5.9} is depicted in Fig. \ref{fig2}.

Let us now turn to the scalar curvature
 $R_1$ and the  four-dimensional effective cosmological constant
 $\Lambda_{eff}=\left.U_{eff}\right|_{extr}$.
The structure of Eqs. \rf{3.4-11-1}, \rf{3.4-11-2} and
\rf{3.4-11-3} suggests to consider $R_1$ and $\Lambda_{eff}$ as
functions of $(z,X,q,\alpha)$.  Eliminating $q$ from  Eqs.
\rf{3.4-11-2}, \rf{3.4-11-3} (with the help of \rf{5.2}) we obtain
\ba{5.10}
R_1&=&\frac{1}{6\alpha}X^{-1/2}(X+4z-1)\; ,\label{5.10-1}\\
\Lambda_{eff}&=&\frac{1}{12\alpha}X^{-1/2}(X-2z-1)\;
.\label{5.10-2}
\ea
The graphics of the functions
\ba{5.11}
R_1(z,X)=0&\Longrightarrow &X|_{(R_1=0)}=1-4z\nn
\\ \Lambda_{eff}(z,X)=0&\Longrightarrow &X|_{(\Lambda_{eff}=0)}=1+2z
\ea
 are included in Fig.
\ref{fig1}. For completeness, we map them also on the
$(z,q)-$plane. Following the same scheme as above we obtain
\ba{5.11a}
R_1(z,q)=0&\Longrightarrow & q|_{(R_1=0)}=2z(4z-3)\ , \quad 0\le z<1/4\nn\\
\Lambda_{eff}=0&\Longrightarrow & q|_{(\Lambda_{eff}=0)}=2z \ea
and the correspondences
\be{5.12}
\begin{array}{lcl}R_1>0&\mbox{for} &X>1-4z\; ,
\ \ q>2z(4z-3)\; ,  \\
\Lambda_{eff}>0 &\mbox{for} &X>1+2z\; , \ \  q>2z\; . \end{array}
\ee
{}From Fig. \ref{fig1} and Fig. \ref{fig2} we see that the
nonlinear model with two-dimensional internal space $M_1$ allows
for stable configurations only in the cases
\be{5.13}
\begin{array}{lcl}\Lambda_{eff}\ge 0 & \mbox{for} & R_1>0\; ,\\
\Lambda_{eff}<0 &\mbox{for} & \sign(R_1)=\pm1,0\; . \end{array}
\ee
It contains no stable configurations with an accelerated expansion
of the Universe $(\Lambda_{eff}>0)$ for internal spaces of
negative or vanishing scalar curvature $R_1$.

Finally, we comment on some limiting cases.
\begin{enumerate}
\item[(L.3.1)] $h\to +0$, $q\neq 0$: According to Figs. \ref{fig1}, \ref{fig2} this limit corresponds to
a vanishing form field $z\to+0$, and a stabilization is possible
in the case of $R_1<0$. Furthermore,  for  $z\to 0$ we can
approximate $X=X_1(z,q)\approx(v-1)(1+z/v)$ with
$v:=\sqrt{4+3q}>1$ and the masses of the normal excitation modes
of the coupled $\varphi-\phi-$field system follow from \rf{3.4-8},
\rf{5.4-1} - \rf{5.4-3} as
\ba{5.14}m_1^2&=&\frac{1}{6\alpha}(v-1)^{-1/2}\left[2-v+\frac92 z +{\cal
O}(z^2)\right]\; ,\nn\\
m_2^2&=&\frac{1}{10\alpha}(v-1)^{-1/2}\left[v-\frac{2+v}{2v}z
+{\cal O}(z^2)\right]\; .
\ea
In the special case $z=0$ we completely reproduce our earlier
results \cite{GMZ(PRDa)} on nonlinear stabilized models without
form fields $(d_1=2)$: $m^2_{\varphi}=m_1^2=-U_0(X)$, $m^2_{\phi}=m^2_2=X^{-1/2}(X+1)/(10\alpha)$.
\item[(L.3.2)] $\alpha\to +0,\Lambda_D, h\neq 0$:
For this transition to a linear
model we have as in (L.1.1) \ $U(X)\to \Lambda_D$ as well as a
freezing of the nonlinearity field at $\phi_0\to 0$,  $X\to 1$.
Using the approximation
\be{5.15}
X=X_1(z,q)=1+z+(3q-2z)/4-(3q-2z)^2/64+z^2/4+{\cal O}(\alpha^3)
\ee
we obtain the excitation masses as
\ba{5.16}
m_{\phi}^2\to m_1^2& =  &\frac{\alpha^{-1}-2h}{5}+{\cal O}(\alpha)\to \infty\;
,\nn\\ m_{\varphi}^2\to m_2^2&= &3h-\Lambda_D+{\cal O}(\alpha)>0\;
\ea
so that the freezing is clearly seen from the diverging mass of
the nonlinearity field.  Additionally, we find from \rf{5.10-1},
\rf{5.10-2}
\ba{5.17}
R_1&=&\Lambda_D+3h
-\frac{\alpha}{6}\left[27 (h+\Lambda_D)^2-8h^2\right]
+{\cal O}(\alpha^2)\label{5.17-1}
\\
\Lambda_{eff}&=&\frac{\Lambda_D-h}{2}+\frac{3}{4}\alpha
\left[(h+\Lambda_D)^2-4\Lambda_D^2\right]+{\cal
O}(\alpha^2),\label{5.17-2}
\ea
what in the special case of a vanishing effective cosmological
constant $\Lambda_{eff}=0$ reproduces the results of Ref.
\cite{GZ1} for a linear model with Freund-Rubin form field:
$h=\Lambda_D=R_1/4$, $m_{\varphi}^2=2h$.
\item[(L.3.3)] $\Lambda_D\to 0$, $\alpha,h\neq 0$: In this case we have
$q\to 0$. A substitution of the approximation
$X=X_1(z,q)=z-1+\left[(z-1)^2+3\right]^{1/2}
+3\left[(z-1)^2+3\right]^{-1/2}q/2-{\cal O}(q^2)$
into the Hessian shows that there is no special behavior
of the excitation masses connected with this limit.
\item[(L.3.4)] $\Lambda_D,h\to 0, \alpha\neq 0$:
{}From \rf{5.4-1} it follows in this limit $J_{11}\to 0$, so that
beside a decoupling of the excitations the gravexciton mass
vanishes $m_1^2\to m^2_{\varphi}\to 0$ and the internal space
$M_1$ destabilizes. This is in full agreement with
\cite{GMZ(PRDa)} where a stabilization for $h=0$ requires
$\Lambda_D<0$.
\item[(L.3.5)] $R_1\to 0$: In the limit $R_1\to 0$ one observes a
regular behavior similar to (L.3.3).
 For parameter points near the line
$X_0(z):=X_{(R_1=0)}(z)=1-4z$,  $0\le z<1/4$ we find from Eq.
\rf{5.10-1} $X=X_0(z)+6\alpha R_1 X_0^{1/2}+{\cal O}(\alpha^2
R_1^2)$ so that the Hessian yields excitation masses of the form
$m^2_{1,2}(R_1\approx 0)=m^2_{1,2}(R_1=0,z)+\sigma_{1,2}(z)\alpha
R_1+{\cal O}(\alpha^2 R_1^2)$ with some regular coefficients
$\sigma_{1,2}(z)$ and
 \be{5.18}m^2_{1,2}(R_1=0,z)=\frac{1}{40\alpha}X_0^{-1/2}\left[9-5X_0\mp
 \sqrt{4(3-5X_0)^2+5(X_0-1)^2}\right]\;
 .
 \ee
The masses $m^2_{1,2}(R_1=0,z)$ have finite values except at the
limiting points $X_0(z\to1/4)\to 0$ (or $\phi_0\to -\infty$) and
$X_0(\alpha \to 0)\to 1$ (or $\phi_0 \to 0$) where both or one of
the masses diverge. We see that, with exception of the limiting
point\footnote{According to \rf{5.18}, the limit $X_0(\alpha \neq
0)\to 1$ gives $m_1^2\to m^2_{\varphi}\to 0$, $m_2^2\to
m^2_{\phi}\to 1/(5\alpha)$. On the other hand, $X_0(\alpha \neq
0)\to 1$ implies $z=4\alpha h\to 0$ and according to \rf{5.3} also
$\Lambda_D \to 0$. Thus, the results of (L.3.1) and (L.3.4) can be
used to reproduce the same behavior of the excitation masses via
\rf{5.14}.} $X_0(\alpha\neq 0)\to 1$, there occurs no
destabilization of the internal space $M_1$ for vanishing scalar
curvatures $R_1$. Due to the smooth behavior of the excitation
masses under the transition $R_1\to 0$ we can identify this limit
with a stable decompactification $r_1\to \infty $ of an internal
space $M_1$ with fixed topology. Clearly, in our local approach a
stable decompactified space with $r_1\to \infty$ is
indistinguishable from a stabilized internal space which is
Ricci-flat from the very beginning.
\end{enumerate}



\section{Conclusions and discussion \label{conclu}}
\setcounter{equation}{0}

In the present paper we investigated multidimensional
gravitational models with a non-Einsteinian form of the action. In
particular, we assumed that the action is an arbitrary smooth
function of the scalar curvature $f(R)$. Additionally, the
D-dimensional spacetime was endowed with solitonic form fields of
generalized block-orthogonal Freund-Rubin type. This bulk matter
ansatz leads to a naturally factorized geometry and a spontaneous
compactification can be associated with it. For the considered
models, we concentrated on the stabilization problem for the extra
dimensions. As technique we used a reduction of the nonlinear
gravitational model to a linear one with an additional
self-interacting scalar field (nonlinearity scalar field). The
factorized geometry as well as the generalized Freund-Rubin ansatz
for the solitonic form field allowed for a dimensional reduction
of the considered models and a transition to the Einstein frame.
As result, we obtained an effective four--dimensional model with
nonlinearity scalar field and additional minimally coupled scalar
fields which describe conformal excitations of the scale factors
of the internal space.

A detailed stability analysis was carried out for the three most
simplest configurations of a model with one internal factor space
$M_1$ and a quadratic curvature term: $f(R) = R + \alpha R^2
-2\Lambda _D$, where $\Lambda_D$ plays the role of a
$D-$dimensional bare (bulk) cosmological constant. These three
configurations are characterized respectively by: 1) a vanishing
four-dimensional effective cosmological constant $\Lambda_{eff}$,
2) a traceless form-field EMT, or 3) a
$(d_1=2)-$dimensional internal factor space $M_1$. For all three
configurations, a stabilization of the internal space is only
possible in the case of a non-negative nonlinearity parameter
$\alpha \ge 0$ and a bulk cosmological constant
$\Lambda_D$  restricted by the condition $q\equiv
8\alpha\Lambda_D>-1$. The transition $(\Lambda_D\to 0,h\to 0)$
is connected with a decompactification $(R_1 \to 0, r_1 \to \infty)$
of the internal space $M_1$. At the same time,  it
leads to a flattening of the effective potential in the direction
of the scale factor excitations and, hence, to a destabilization of $M_1$
(for a similar limiting behavior see also Ref. \cite{GMZ(PRDa)}).

{}From the three configurations, the model with the two-dimensional
internal space shows the richest features. It allows for stable
configurations in the cases $(\Lambda_{eff}\ge 0, R_1>0)$ and
$(\Lambda_{eff}<0, \mbox{any sign of } R_1)$ as well as for
Ricci-flat internal spaces  $(R_1=0)$. Interestingly, the various
stable configurations belong to a connected region in the
parameter space ${\cal M}$ and one can smoothly pass from one type
of configuration to another one, including a transition to stable
Ricci-flat internal spaces which can be described as "stable
decompactifications": $R_1 \to 0$, $r_1\to \infty$. As pointed out
in the Introduction, such a rich picture became possible due to
the presence of the real-valued form fields which satisfy the NEC
and the WEC and which compete with the nonlinearity scalar field.
The latter satisfies the NEC only marginally and can violate the
WEC.

Interestingly, for  $(d_1=2)-$dimensional internal spaces there
exist  parameter configurations with $\alpha,\Lambda_D,h,R_1>0$
that can provide positive values of the effective four-dimensional
cosmological constant $\Lambda_{eff}> 0$ (see e.g. Eqs. \rf{5.12},
\rf{5.13}). Thus, an accelerated expansion of the Universe seems
possible in accordance with observational data. Let us assume that
the values of the bulk cosmological constant $\Lambda_D$ and the
form field strength $h$ are set at some characteristic scale
$\Lambda_D\sim h\sim \bar{M}^2$.  Then we find for the parameters
$q\sim 8\alpha \bar{M}^2 $, \ $z\sim 4\alpha \bar{M}^2 $ and,
hence, $q\sim 2z$. The latter corresponds to $X\sim 1+2z\sim
1+8\alpha \bar{M}^2 $ (see \rf{5.3}) and comparison with
\rf{5.11a} shows that such configurations should yield an almost
vanishing effective cosmological constant $\Lambda_{eff}\sim 0$.
With the help of Eq. \rf{5.10-1} we can estimate the scalar
curvature $R_1$ of the internal space as
\be{6.0}
R_1\sim \frac{z}{\alpha}X^{-1/2} \sim
\frac{1}{\alpha}\frac{z}{\sqrt{1+2z}} \sim \frac{4\bar{M}^2
}{\sqrt{1+8\alpha \bar{M}^2 }}\; .
\ee
On the other hand, its value is connected with the fundamental
  scales $M_{*(4+d_1)} $, $M_{Pl(4)} $  by the relations \rf{0.1}, \rf{0.2}
and \rf{2.5a}:
\be{6.1}
R_1\sim r_1^{-2}\sim
\left(\frac{M_{*(4+d_1)}}{M_{Pl(4)}}\right)^{4/d_1}M_{*(4+d_1)}^2\; .
\ee
As mentioned in the discussion after Eq. \rf{1.28}, the value of
$X$ can be used as a measure of the nonlinearity of the original
model: $\alpha \ov R=e^{A\phi_0}-1\equiv X-1$. We see that weakly
nonlinear configurations correspond to $X\approx 1$, whereas $X\gg
1$ leads to a strongly nonlinear regime. With the help of \rf{6.0}
and \rf{6.1} we express this dimensionless nonlinearity parameter
$X$ in terms of the different scales contained in
 our model:
\be{6.2}X\sim 1+8\alpha \bar{M}^2
\sim 16 \left(\frac{\bar M}{M_{*(4+d_1)}}\right)^4
\left(\frac{M_{Pl(4)} }{M_{*(4+d_1)} }\right)^{8/d_1}\; .
\ee
{}From \rf{6.2} we see that
setting  $\bar M \sim M_{*(4+d_1)} $ we obtain
$X\gg 1$ for ADD-type TeV$-$scale models
whereas $X\sim 1$ can only be achieved for standard KK models with
$ M_{Pl(4)} \sim M_{*(4+d_1)} $.
 Stably compactified internal spaces in ADD-type models can be obtained
within a weakly nonlinear regime $X\sim 1$ if
 the bulk cosmological constant
$\Lambda_D$ and the form field strength $h$ are related with the fundamental
scales as
\be{6.3}
\Lambda_D \sim h\sim \bar{M}^2, \qquad \bar M
\sim \frac 12 M_{*(4+d_1)}
\left(\frac{M_{*(4+d_1)} }{M_{Pl(4)} }\right)^{2/d_1}.
\ee
For $M_{*(4+d_1)}\sim 1 - 30 $ TeV and $d_1=2$ this implies $\bar
M \sim 10^{-4} - 10^{-1}$eV. It is interesting to note that this
mass scale is of the same order as the lowest possible
supersymmetry breaking scale $m\sim M^2_{SUSY}/M_{Pl(4)} $ in the
minimal supersymmetric extension of the standard model (MSSM)
\cite{susy} with $M_{SUSY}\sim 1$ TeV.

Above we demonstrated that the assumption $\Lambda_D\sim h$ can
result in a small effective cosmological constant $\Lambda_{eff}$.
Let us now estimate the relation between $\Lambda_{eff}$ and
$\Lambda_D, h$ in more detail and compare it with the observable
value\footnote{In our normalization conventions holds $c=\hbar =1$
and $\Lambda_{Pl(4)}\sim M^2_{Pl(4)}\sim L^{-2}_{Pl(4)}$.} of
$\Lambda_{eff}\sim 10^{-123}\Lambda_{Pl(4)}$. For simplicity, we
will restrict our consideration to a weak nonlinearity regime with
$X\approx 1$, $\alpha  \gtrsim 0 $ where the approximations
\rf{5.17-1} and \rf{5.17-2} of (L.3.2) are valid.
 {}From \rf{5.17-2} we see that to ensure  a
sufficiently small $\Lambda_{eff}$ the bulk cosmological constant
$\Lambda_D$ and the field strength $h$ of the solitonic form field
should be connected by
\be{6.6}
h=(1+\epsilon)\Lambda_D\; .
\ee
The value of the small $\epsilon$ we will estimate now.  With the help
of relations \rf{5.17-1} and \rf{6.3} we find
\ba{6.7}
10^{-123}\Lambda_{Pl(4)}\sim \Lambda_{eff}&\approx &\epsilon
\Lambda_D(1-6\alpha \Lambda_D)/2
\\ \Lambda_D &\sim &  \bar{M}^2 \sim R_1 \nn
\ea
and, hence,
\be{6.8}
10^{-123}\sim \epsilon
\left(\frac{M_{*(4+d_1)}}{M_{Pl(4)}}\right)^{4/d_1+2}
\ee
so that
\be{6.9}
\epsilon \sim 10^{-65}
\ee
for $d_1=2$, $M_{*(4+d_1)}\sim 30$TeV. According to \rf{6.7} this
value of $\epsilon$ is not sensitive to changes of the
nonlinearity parameter $\alpha$ in a weakly nonlinear curvature
regime. Thus, we arrive at the conclusion that the ADD scenario in
its simplest extended version can provide a simultaneous
stabilization of the extra dimensions together with an adjustment
of the effective cosmological constant to its observed value only
in the case of a strong fine tuning. Although the solitonic form
fields of our model are located in the compactified extra
dimensions, the tuning of their effective energy density  $h$ to
the bulk cosmological constant $\Lambda_D$: \
$h=(1+\epsilon)\Lambda_D$ is of a similar type as the
four-form-tuning discussed in Weinberg's no-go theorem \cite{ccp1}
for a resolution of the cosmological constant problem (CCP). A
shifting of the CCP to a parameter fine tuning is a rather general
feature of models with compactified additional dimensions and form
fields\footnote{For a similar mechanism in RSII-type models with
form fields see Ref. \cite{BG1}.}. In a slightly reshaped form it
also appears in the recently proposed brane-world model with
two-dimensional "football"-shaped large extra dimensions
\cite{ccp2} (see also \cite{navarro}) where the adjustment of the
on-brane cosmological constant is shifted to an adjustment of the
parameters of the off-brane "football". A possible resolution of
the CCP for similar higher dimensional models with form fields
following from an M-theory setup was presented in Ref.
\cite{ccp3}. Proposals for a resolution of the CCP within other
scenarios comprise various anthropic approaches
\cite{BG1,anthrop}, shifting of the CCP to a singularity problem
\cite{sing1}, possible graviton compositeness \cite{sundrum2}, a
holographic approach \cite{banks} as well as non-local
modifications of gravity \cite{dvali10}. However, there is still
no satisfactory and comprehensive solution of the CCP. The problem
will probably remain challenging the scientific community until a
final understanding of quantum gravity will be achieved.


\bigskip
{\bf Acknowledgments}

U.G. and A.Z. thank H. Nicolai and the Albert Einstein Institute,
as well as the Department of Physics of the University of Beira
Interior for their kind hospitality during the preparation of this
paper. The work of A.Z. was supported by a BCC grant from
CENTRA--IST  and partly supported by the programme SCOPES
(Scientific co-operation between Eastern Europe and Switzerland)
of the Swiss National Science Foundation, project No. 7SUPJ062239.
U.G. acknowledges support from DFG grant KON/1344/2001/GU/522.
Additionally, this research work was partially supported by the
grants POCTI/32327/P/FIS/2000, CERN/P/FIS/43717/2001 and
CERN/P/FIS/43737/2001.


\end{document}